\documentclass[12pt]{article}
\usepackage{amsmath}
\usepackage{txfonts}

\usepackage{graphicx}

\usepackage[letterpaper,margin=1in]{geometry}

\usepackage{parskip} 

\renewenvironment{abstract}
	{\quotation}
	{\endquotation}

\date{}

\makeatletter
\renewcommand{\fnum@figure}{\textbf{Figure \thefigure}}
\renewcommand{\fnum@table}{\textbf{Table \thetable}}

\makeatother

\usepackage{url}

\newcommand{\pdag}{{\phantom{\dag}}}

\def\scititle{
Negative differential conductance in triangular molecular assemblies
}
\title{\bfseries \boldmath \scititle}

\author{
    Chao Li$^{1,2\dagger\ast}$,
    Vladislav Pokorn\'y$^{3\dagger\ast}$,
    Prokop Hapala$^{3\dagger}$,
    Martin \v{Z}onda$^{4}$,
    Ping Zhou$^{5}$,
    \and
    Silvio Decurtins$^{5}$, 
    Shi-Xia Liu$^{5\ast}$,
    Fengqi Song$^{1}$,
    R\'emy Pawlak$^{2}$,
    Ernst Meyer$^{2}$
    \and
    \small$^{1}$Institute of Atom Manufacturing, Nanjing University, Suzhou 215163, China. \and
    \small$^{2}$Department of Physics, University of Basel, Klingelbergstrasse 82, 4056 Basel, Switzerland.\and
    \small$^{3}$Institute of Physics (FZU), Czech Academy of Sciences, Na Slovance 2, 182 00 Prague 8, Czech Republic.\and
    \small$^{4}$Department of Condensed Matter Physics, Faculty of Mathematics and Physics, Charles University,\\
    \small Ke Karlovu 5, 121 16  Prague 2, Czech Republic.\and
    \small$^{5}$Department of Chemistry, Biochemistry and Pharmaceutical Sciences, W. In\"abnit Laboratory \and
    \small for molecular quantum materials and WSS Research Centre for Molecular Quantum Systems, \and
    \small University of Bern, Freiestrasse 3, 3012 Bern, Switzerland.
    \and
    \small$^\ast$Corresponding author. Email: chao.li@nju.edu.cn; pokornyv@fzu.cz; shi-xia.liu@unibe.ch;
    \and
    \small$^\dagger$These authors contributed equally to this work.
}

\begin{document} 
\maketitle

\begin{abstract} \bfseries \boldmath
We report the creation and characterization of a molecular-scale negative differential conductance (NDC) device by assembling a triangular trimer of 4,5,9,10-tetrabromo-1,3,6,8-tetraazapyrene (TBTAP) molecules on a superconducting Pb(111) substrate. Using low-temperature scanning tunneling spectroscopy, we observe robust NDC behavior manifesting as a decrease in current with increasing voltage between 0.7–0.9~V arising from the interplay of Coulomb blockade and strong inter-molecular capacitive coupling within the molecular cluster. Gate-controlled charging and discharging processes are directly visualized via two-dimensional differential conductance mapping, which reveals the emergence of Coulomb rings and spatial regions of NDC. Theoretical modeling using a three-impurity Anderson model and master equation approach quantitatively reproduces the experimental observations and demonstrates that the NDC emerges purely from electron correlations, independent of the underlying superconductivity. By tuning the geometry to a hexamer structure, we further show that cluster topology provides versatile control over electronic properties at the molecular scale. These results establish a functional platform for implementing multifunctional molecular devices and highlight a strategy toward programmable and scalable nanoelectronics.
\end{abstract}

\newpage
The relentless pursuit of miniaturization in electronics has driven research toward nanoscopic devices, where quantum phenomena and atomistic details play crucial roles. While these phenomena complicate the fabrication of traditional semiconductor devices, they can also be harnessed to implement nontrivial functionalities in molecular-scale systems, achieving ultimate miniaturization~\cite{Xiang-2016,Evers-2020,Famili-2019}. Unlike semiconductor junctions and quantum dots, identical molecules can be mass-produced with atomic precision via environmentally friendly and energy-efficient methods of organic synthesis. Due to their sub-nanometer size, molecular components can efficiently exploit effects such as coherent resonant transport, Coulomb blockade, and strong inter-molecular capacitive coupling, phenomena essential for implementing nontrivial functions, including memories, memristors, switches, and oscillators. 

Among such behaviors, negative differential conductance (NDC), where increasing voltage results in decreasing current, is of particular importance. NDC enables high-speed switching, low-power logic, and oscillatory circuits, and has become a fundamental mechanism for emerging nanoelectronics, molecular computing and quantum sensing~\cite{Bedrossian-1989, Lyo-1989,Xu-2015,Nowakowski-2025-arXiv}. While NDC has been well-established in conventional semiconductor components (e.g., Gunn diodes, Esaki diodes)~\cite{Knight-1967-GunnTheory,Esaki-1966}, quantum dot arrays~\cite{Rogge-2006,Kostyrko-2009,Weymann-2018}, two-dimensional heterostructures~\cite{Kim-2013,He-2017,Zhao-2021}, molecular layers~\cite{Grobis-2005-C60layers,Chen-2007,Torrente-2012} and even single-molecule break junctions~\cite{Heersche-2006,Perrin-2014}, scalable integration of NDC into reproducible molecular circuits has been hampered for decades by the so-called wiring problem. That is, the characteristics of molecular devices connected via metallic leads are extremely sensitive to atomic-scale details of the metal–molecule interface, which lie beyond the control of standard nanofabrication methods, making consistent device fabrication exceedingly difficult.

To avoid this bottleneck, recent nanoelectronics research has shifted toward alternative architectures, notably quantum molecular cellular automata assembled on atomically flat substrates \cite{Lent2003,Bandyopadhyay2010,Berger2020}. In these systems, molecular components interact via local through-space capacitive couplings or spin-mediated interactions, eliminating the need for wired connections~\cite{li2017conformation}. In our previous works, we demonstrated that 4,5,9,10-tetrabromo-1,3,6,8-tetraazapyrene (TBTAP) molecules exist as radicals when adsorbed on both Ag(111) and Pb(111) surfaces\cite{li2023strong,pawlak-2023-arXiv}, making them an ideal platform for investigating electron-electron interactions within designed structures\cite{Li-2025}. As an example, an assembly of four TBTAP molecules on superconducting Pb(111) surface exhibits robust switching behavior due to strong capacitive coupling, effectively implementing a molecular memory device~\cite{Li-2025}.

In this work, we report the first observation of NDC in an molecular system engineered using lateral manipulation, specifically, in a $C_{3}$-symmetric triangular trimer of the same molecules on a Pb(111) substrate, forming a compact functional unit of nanometer size, measured using low-temperature scanning tunneling spectroscopy (STS). The NDC manifests itself in the differential conductance (d\textit{I}/d\textit{V}) spectra at voltages between 0.7-0.9~V and originates from the Coulomb blockade effect, driven by the discrete charging energies of the cluster. By adjusting the gate voltage, electrons can be discharged from the molecules, as evidenced by the appearance of Coulomb rings in d\textit{I}/d\textit{V} mapping (Fig.~\ref{Figure1}A).  When the Coulomb rings of neighboring molecules overlap, we observed distinct features, including the emergence of regions of NDC. Although the measurements were performed on a superconducting substrate, the NDC behavior is unrelated to superconductivity and instead arises purely from correlated hopping of electrons between the scanning tunneling microscope (STM) tip and the molecules, governed by the inter-molecular Coulomb correlations. Expanding the trimer to a hexamer with the same $C_{3}$ symmetry introduces additional complexity as the three inner molecules exhibit Coulomb rings while the states of the three outer molecules remain passive to the presence of the charged scanning tip. This shows how the geometry of the cluster can be further exploited to fine-tune the functionality of such molecular assemblies.
 
Unlike traditional semiconductor-based NDC elements such as resonant tunneling diodes, molecular systems offer not only unmatched spatial compactness, but also seamless integration with quantum and neuromorphic computing frameworks \cite{Wasielewski2020}. Furthermore, our findings underscore the role of a molecular cluster topology: switching between linear and triangular configuration yields distinct device behaviors - molecular memory~\cite{Li-2025} vs. molecular Gunn diode - without altering the molecular building blocks. This demonstrates a powerful approach to diversifying functionality within a single chemical system.

The same system also exhibits Yu–Shiba–Rusinov (YSR) states inside the superconducting gap at low bias voltage, potentially enabling the creation of superconducting quantum devices, such as functionalized nanoscopic Josephson junctions, within the same molecular platform. This opens a promising route toward hybrid devices that integrate multiple quantum functionalities, including superconductive electronics, a leading beyond-CMOS platform for future energy-efficient computing~\cite{Bairamkulov-2024SCElectronics}.

Altogether, our results highlight the potential of rationally designed molecular clusters for enabling complex, nonlinear behavior at the ultimate limit of miniaturization. Controlled NDC at the molecular scale offers a pathway to realize novel components such as single-electron transistors, molecular oscillators, and logic gates for quantum-inspired computing. By bridging fundamental quantum effects with practical functionality, this work marks a significant step toward the realization of programmable, self-assembled nanoelectronics.

\subsection*{Triangular trimer assemblies and their discharging behavior}
We achieved a precise assembly of triangular molecular trimers with $C_{3}$ symmetry by manipulating TBTAP molecules on the Pb(111) surface using a lateral manipulation technique (Fig.~\ref{FigSM:lateral}). The molecules form an equilateral triangle with a 1 nm distance between the centers of the molecules (Fig.~\ref{Figure1}B). The trimer is stabilized through a combination of hydrogen bonding and $\text{N}\cdots\text{Br}$ halogen bonding, in agreement with a DFT simulation (Fig.~\ref{FigSM:DFT}), creating a chiral structure. The chirality in this case is the result of incommensurate size of the TBTAP molecules with respect to the surface lattice constant of Pb(111). 
Tunneling (d\textit{I}/d\textit{V}) spectra and current were recorded at positions marked by colored dots in Fig.~\ref{Figure1}B at temperature $T_\text{exp}=2.6$~K. As shown in Fig.~\ref{Figure1}C, the d\textit{I}/d\textit{V} spectra exhibit sharp peaks at $V_\text{s}=520$ mV on individual molecules (yellow, red and blue lines) and at $V_\text{s}=590$ mV at the center of the trimer (green line). 
The discharging peaks gradually shift toward the Fermi energy as the tip height $z_{\text{tip}}$ decreases (Fig.~\ref{FigSM:differentheight}), consistent with the tip-induced electric field effect~\cite{wu2004control,kumar2019electric,li2023strong}. In addition to these sharp peaks, satellite peaks and NDC regions, characterized by dips in the spectra, were observed at energies close to $V_\text{s}=760$ mV. These regions are marked by drops in current with increasing voltage, as shown in Fig.~\ref{Figure1}D, showing that NDC is a genuine effect. 

Fig.~\ref{Figure1}E displays the position-dependent d\textit{I}/d\textit{V} spectra along the red line indicated in Fig.~\ref{Figure1}B, which connects the centers of two molecules. Similarly, Fig.~\ref{Figure1}F displays the spectra measured along the green line that connects the center of the trimer to the center of a molecule. Data show bands of high positive and high negative conductance, which correspond to jumps in the tip current, which are usually attributed to a gain or loss of an electron at the given molecule. 

Two-dimensional STS mapping of the TBTAP trimer provides a more direct insight into the discharge behavior.
Fig.~\ref{Figure2}A illustrates a two-dimensional d\textit{I}/d\textit{V} map of the trimer at $V_\text{s}=540$ mV, revealing the presence of three distinct discharge rings corresponding to the three individual molecules (see also Fig.~\ref{FigSM:superposition}). Similar rings were previously observed in different nanostructures~\cite{Pradhan-2005,Nazin-2005,Wong-2015,Kocic-2015,Li-2021,pawlak-2023-arXiv}. The evolution of discharging rings with increasing $V_\text{s}$ from $570$ to $870$ mV is illustrated in Figs.~\ref{Figure2}B-H. As $V_\text{s}$ increases, the rings expand in size and begin to overlap at $V_\text{s}=570$ mV, leading to a flower-like patterns. Upon further increase in the voltage, regions of NDC, holding a distinct chiral pattern, emerge near the center of the trimer. Similar flower-like characteristics were previously observed in WS$_2$/WSe$_2$ moir\'e lattices, though without any indication of NDC regions~\cite{Li-2021}.

In contrast to the extended layered systems as discussed in, e.g.,  Refs.~\cite{Torrente-2012,Li-2021}, the finite size of our system allows for full theoretical description using a three-impurity Anderson model (TIAM). Details of the method and its implementation are summarized in the Supplementary Material. The molecules are represented as single correlated quantum levels at energy of $-90$~meV with respect to the surface chemical potential that interact via inter-site Coulomb (capacitive) coupling. The low-energy equilibrium physics of TIAM can be addressed by the numerical renormalization group (NRG) method. However, to gain insight into the behavior at sizable voltage, we performed a simulation utilizing the Pauli master equation (PME) approach. Here we assume that the molecules are weakly coupled to the tip and the substrate ($\Gamma_\text{t}<\Gamma_\text{s}\ll k_BT_\text{exp}=0.22$~meV). Although this assumption might be questionable, since the value of the substrate coupling, fitted from the YSR characteristics~\cite{Li-2025}, is $\Gamma_\text{s}=10-20$~meV, the size of the discharge rings depends mainly on the local electrostatic environment and not on the coupling. Such a small value of the coupling puts the system in the sequential tunneling regime, greatly simplifying the analysis of the experiment, as we can neglect any cotunneling effects. The on-site Coulomb interaction strength $U=200$~meV 
is large enough to prohibit double occupancy of a site, significantly reducing the size of the configuration space of the molecular subsystem from $4^3=64$ to $2^3=8$ states. Other parameters of the model were fitted from a broad set of experimental data plotted in Fig.~\ref{FigSM:maps_dIdV}. The results of the simulations are presented in Figs.~\ref{Figure2}A'-H'. The spatial distribution of the electron density associated with the SOMO in the TBTAP trimer was calculated using DFT (Fig.~\ref{FigSM:DFT}) and incorporated into the simulation to elucidate the effects of the chirality, although it is not essential for the description of NDC. The structure of the SOMO orbital was omitted for data presented in Fig.~\ref{FigSM:maps_dIdV} and replaced by a simple s-orbital, while all the main features, except the chiral pattern, are still present, showing the robustness of the effect. The slightly oval shape of the discharge rings in the experiment was introduced into the simulation by a finite quadrupole moment of the molecules. The almost perfect match between the experiment and the theory allows us to pinpoint physical parameters such as 
the value of the inter-site Coulomb coupling $W=50$~meV, which is consistent with previous studies of TBTAP dimers~\cite{Li-2025}. The simulation also suggests a negligible value of the direct hopping between orbitals. As a result, charge rearrangement within the cluster is allowed only via the conductive substrate.

Low-energy spectra were also measured and show two pairs of well-developed YSR states inside the superconducting gap at $V_\text{s}=\pm1.85$~mV and $1.95$~mV (Fig.~\ref{FigSM:YSR}). The spatial d\textit{I}/d\textit{V} maps measured at this energy again show a chiral pattern originating from the molecules' relative orientation. The YSR spectra were calculated using the NRG solution of the superconducting TIAM for the model parameters inferred from the simulations of the charging rings. The results show doubly-occupied ground state and a pair of slightly split YSR peaks, in agreement with the experiment, providing additional check for the reliability of the model calculations. Additional details are provided in the Supplementary Material.

\subsection*{Negative differential conductance emerging from non-equilibrium occupancy of states}
To understand the origin of the NDC observed in Fig.~\ref{Figure2}, we analyzed the individual components of the tunneling current obtained from PME calculations. The components, as well as their correspondence to the many-body state energies are shown in Figs.~\ref{Figure3}A-B. Data are plotted for sample voltage $V_\text{s}=790$~meV, along the line represented by red arrows in Figs. 3C-E, which passes directly over molecule 1 and between molecules 2 and 3. To keep the explanation simple, we assume a zero value of the quadrupole moment, which otherwise lifts the degeneracy of the on-site energies of molecules 2 and 3 (Fig.~\ref{FigSM:mbe}B). The key insight lies in the significant capacitive coupling $W\sim 50$~meV, which makes charging the cluster with two electrons to $-2e$ more energetically favorable than charging it with three electrons to $-3e$, despite the on-site energies being located $90$~meV below the Fermi level of the substrate (i.e., the molecular LUMO being charged into SOMO). This results in a frustrated charge state, where typically two of the three molecular sites are occupied and one remains empty. As the direct hopping between the sites is negligible, we consider the electrons to be localized on individual sites. Upon applying a sufficiently high bias voltage (Sample voltage $V_\text{s}=0.6$~V when tip is at the center of the cluster), an additional electron is repelled from the cluster, leading to a net charge of $-1e$, meaning that only one site remains occupied. 

The total charge of the cluster for different regions is marked in Figure~\ref{Figure3}D, which shows the two-dimensional map of d\textit{I}/d\textit{V}, that corresponds to the experimental data plotted in Fig.~\ref{Figure1}F. In contrast to the intuitive interpretation of discharge rings in the d\textit{I}/d\textit{V} maps as indicating changes in total charge, many of the sharp transitions in site occupancy (manifesting as peaks in d\textit{I}/d\textit{V} spectra) actually correspond to intra-cluster charge rearrangements, i.e., electrons shifting between different molecular sites (through the conductive substrate) without change of total charge. This can be understood by analyzing the many-body ground state energies plotted in Fig.~\ref{Figure3}A. In a quasi-equilibrium regime (i.e., when the current from the tip is negligible and does not perturb the site occupancy), the system behavior is governed by the ground state alone. As the tip position varies, the locally generated repulsive electric field destabilizes nearby occupied sites. In the language of many-body states, this causes the ground state to switch to a configuration where the site directly beneath the tip is empty, opening up a tunneling channel from the tip to that site.

However, as illustrated in Fig.~\ref{Figure3}B, these ground-state transitions only correspond to positive peaks in the d\textit{I}/d\textit{V} spectrum, as expected. The NDC arises instead in the central region between the excited-state crossings, where the energies of singly occupied excited states $|100\rangle$, $|010\rangle$, and $|001\rangle$ become lower than doubly occupied excited states (e.g., $|101\rangle$ and $|110\rangle$). Due to the absence of inter-site hopping in our model, these singly occupied states cannot transit from one state to another by single-electron tunneling processes, leading to non-equilibrium occupancy distributions. Consequently, the system enters a mixed state in which the occupancy is approximately evenly distributed between the sites. 

The resulting suppression of conductance in this central region can be understood in terms of distance dependence of tunneling between the tip and sites, which is in our simplified case modeled as $t_{ti} \propto \exp(-\beta |\mathbf{r}_i - \mathbf{r}_t|)$, where $\mathbf{r}_i$ is site position, $\mathbf{r}_t$ tip position and $\beta$ is a decay constant. In contrast to the quasi-equilibrium regime, where the tip field pushes the electron out of the nearest site, which also has the highest tunneling amplitude, in the non-equilibrium regime, a large fraction of the occupation probability is "wasted" in states where the unoccupied site is far from the tip, leading to ineffective tunneling. In reality, the angular dependence of tunneling matrix elements considering symmetry of the relevant molecular orbital suppresses the current for some tip positions due to nodes of the orbitals, reducing the NDC region and giving rise to the chiral pattern observed in both experiment and simulation (Fig.~\ref{Figure2}).

\subsection*{Negative differential conductance in triangular hexamer assemblies}
Another type of triangular structure, composed of six TBTAP molecules, was constructed using the same lateral manipulation technique (for details on the geometry see Fig.~\ref{FigSM:hedidv}A). The assembly can be manipulated across the Pb(111) surface, showing high structural stability (Fig.~\ref{FigSM:hexla}). The hexamer structure introduces two distinct molecular environments. The three inner TBTAP molecules at the core of the assembly form a trimer similar to the one discussed in the previous section; however, the three outer molecules modify its local electrostatic environment. Fig.~\ref{FigSM:hedidv}B displays the d\textit{I}/d\textit{V} spectra acquired from the inner and outer TBTAP molecules, marked by colored dots. In particular, charging dips are observed for the three inner molecules for sample voltages between -400 and -500~mV, whereas the outer molecules do not display significant features within this voltage range. The spectra suggest that the inner molecules are initially at $V_\text{s}=0$ in neutral TBTAP$^0$ state, while the outer molecules are charged. 
This is in agreement with our previous results on short chains of TBTAP molecules, which show that the molecule can lose its radical nature due to the presence of additional electrostatic field from neighboring molecules~\cite{Li-2025}.
In addition, YSR spectroscopy in Fig.~\ref{FigSM:hedidv}D further proves the presence of these two charge states in the hexamer (for details, see the Supplementary Material).

Figs.~\ref{Figure4}A-I illustrate the evolution of the charging rings in d\textit{I}/d\textit{V} spectra with $V_\text{s}$ decreasing from $-0.37$ V to $-0.64$~V (the corresponding current maps are presented in Fig.~\ref{FigSM:hex-current}). The charging rings around the inner molecules in the hexamer increase in size and intersect at $V_\text{s}=0.45$~V. The rings show a distinct four-lobe pattern, similar to the trimer case, as a result of the TBTAP orbital structure. Furthermore, just before the rings intersect, a chiral pattern of positive d\textit{I}/d\textit{V} emerges near the center of the structure for sample voltages around $-0.43$~V. However, in contrast to the trimer, no new patterns emerge at the center of the structure after the rings intersect.

We did not perform a thorough theoretical analysis of the hexamer system, as the behavior is more monotonous and less surprising in comparison to the trimer. The additional electrostatic field from the charged outer molecules shifts the local energy level of the inner molecules above the substrate chemical potential, leaving them unoccupied at zero bias. At negative sample voltage (i.e., positive tip voltage), the electric field of the tip pushes these states below the substrate's Fermi level, causing them to become occupied. This event gives rise to the charging rings observed around each inner molecule, which exhibit NDC at all voltages beyond the charging threshold. The drop in STM current can be understood within the framework observed in TCNQ layers~\cite{Torrente-2012}. On one hand, the charging of an inner molecule opens a new, sequential tunneling channel via its now-occupied orbital. On the other hand, the local negative charge simultaneously increases the tunneling barrier for all tunneling paths (i.e. non-resonant current from the substrate, and from the outer molecules). The net effect is a decrease in total conductance. Besides the charging rings observed over the inner molecules, one would also expect discharging rings to form around the outer molecules at a positive sample voltage. This was not observed in the experiment as the voltage needed to discharge the outer molecules might be larger than 1.2~V, which could damage the sample.

\subsection*{Conclusions and outlook}
We have demonstrated that a well-defined cluster of three molecules, assembled with atomic precision using lateral manipulation, exhibits both positive and negative differential conductance at positive sample voltages arising from correlated intra-cluster charge dynamics. Thanks to the finite and tractable Hilbert space of the molecular subsystem (8 states), we are able to fully simulate its non-equilibrium dynamics and rationalize the observed conductance features using PME approach. This stands in contrast to more complex systems, such as extended molecular layers, where modeling becomes intractable due to the exponential growth of many-body states and the need to account for inter-site hybridization.

Our analysis reveals that the prominent conductance peaks observed in d\textit{I}/d\textit{V} spectra are not necessarily associated with changes in the total charge of the cluster, but often originate from charge rearrangements between molecular sites while the total charge remains conserved. In particular, we identify a regime where the non-equilibrium occupation of excited states leads to a suppression of tunneling current, resulting in NDC. This effect can be understood by considering both the repulsive electrostatic potential and the exponential decay of tunneling amplitudes with distance, leading to reduced current when the probability distribution of charges delocalizes over all three sites. These effects are further modulated by orbital symmetry, giving rise to the observed chiral patterns. 

The observation of similar phenomena in a more complex hexamer structure further confirms the robustness of these effects. The ability to construct such molecular devices with atomic-scale precision and fully understand their electronic behavior opens a path toward rational design of molecular logic architectures. While the present system was created using low-temperature SPM manipulation, future efforts may aim to scalable realization of such functional clusters through templated self-assembly~\cite{Manikandan-2024}, allowing eventual mass production. Our findings thus mark an important step toward the long-term vision of molecular electronics based on programmable arrays of interacting charge states, conceptually akin to classical cellular automata, but operating at the ultimate limit of miniaturization.

\begin{figure}
	\centering
	\includegraphics[width=0.9\textwidth]{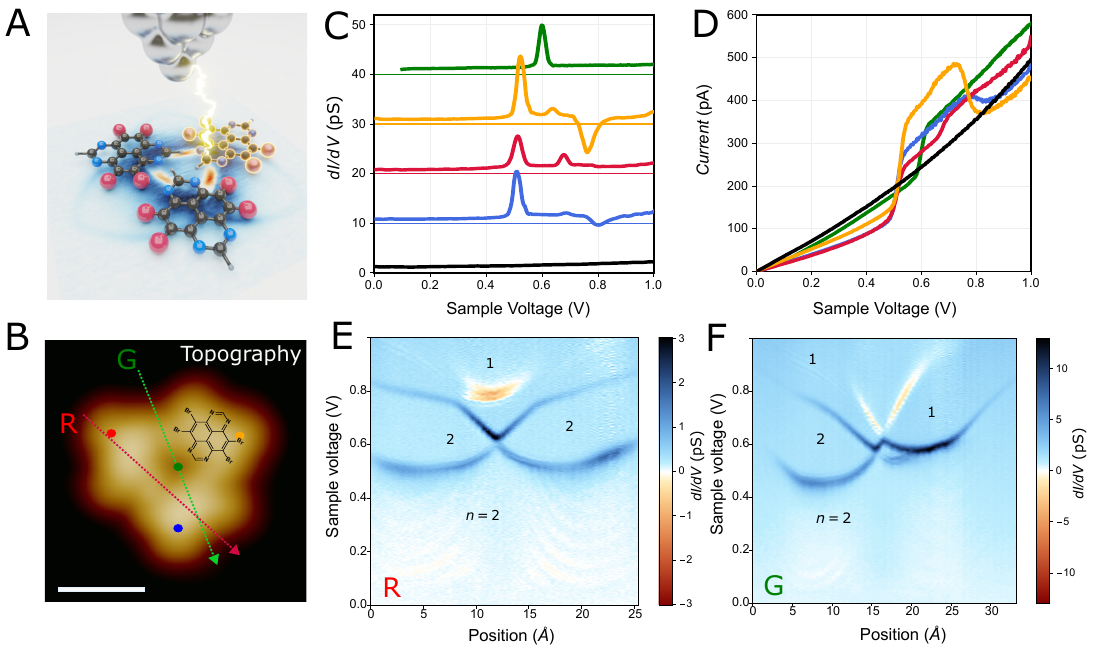}
	\caption{\textbf{Triangular trimer structures of TBTAP molecules.}
		(\textbf{A}) A schematic diagram showing the design of the molecular trimer structure. 
        The local electric field of an STM tip can discharge anionic TBTAP molecules (blue) to neutral (orange), leading to observation of Coulomb rings in d\textit{I}/d\textit{V} maps (blue). The molecules are coupled via inter-site Coulomb interactions. 
        (\textbf{B}) High-resolution STM image ($I$= 100 pA, $V_{\text{s}}$= 600 mV) of a triangular structure constructed from three TBTAP molecules. The scale bar is 1~nm. 
        (\textbf{C} and \textbf{D}) Tunneling spectra (C) and current (D) measured at positions given by color points in panel B. The black lines represents the spectrum and current measured over a pure Pb surface. Spectra are vertically shifted for clarity.
        (\textbf{E}) Position-dependent d\textit{I}/d\textit{V} spectra along the red line shown in panel B, which passes over the centers of two molecules. The total number of electrons sequentially decreases from two to one with increasing sample voltage, as inferred from the simulations. Note that not all peaks in d\textit{I}/d\textit{V} correspond to a change of the total charge on the trimer.
        (\textbf{F})
        Position-dependent d\textit{I}/d\textit{V} spectra along the green line shown in panel B, which connects the center of the trimer and a center of a molecule. 
        }
	\label{Figure1} 
\end{figure}

\begin{figure}
	\centering
	\includegraphics[width=0.9\textwidth]{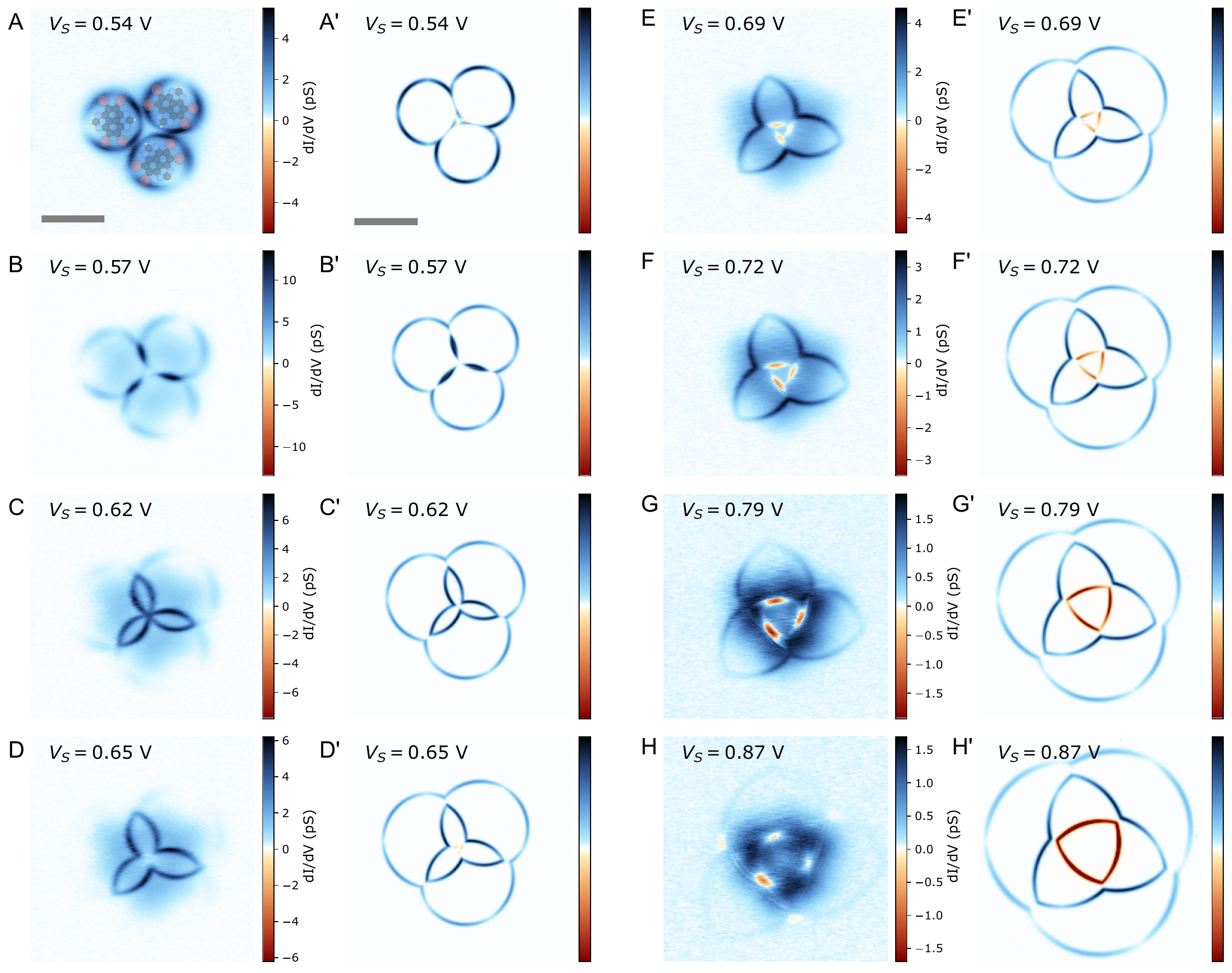}
	\caption{\textbf{Discharging rings in a molecular trimer structure.}
	 (\textbf{A-H}) 
     Evolution of discharging rings with increasing sample voltage $V_\text{s}$ for various voltages: $V_\text{s} = 0.54~\text{V}$ (A), $0.57~\text{V}$ (B), $0.62~\text{V}$ (C), $0.65~\text{V}$ (D), $0.69~\text{V}$ (E), $0.72~\text{V}$ (F), $0.79~\text{V}$ (G), and $0.87~\text{V}$ (H). As $V_\text{s}$ increases, the rings expand and intersect, and regions of NDR with distinct chiral pattern emerge in the center of the trimer. Panel A includes an overlaid molecular structure to indicate the positions of the molecules in the trimer. Br atoms are marked in red, N in blue, C in gray, and H in white. (\textbf{A'-H'}) The evolution of the discharge rings calculated using PME approach (in arb. units) for same sample voltages $V_\text{s}$ as in the experiment. The effects of the TBTAP orbital structure were included in the calculation to simulate the chiral pattern in the experimental data. The scale bars are 1~nm.
     }
	\label{Figure2}
\end{figure}

\begin{figure}
	\centering
    \includegraphics[width=0.9\textwidth]{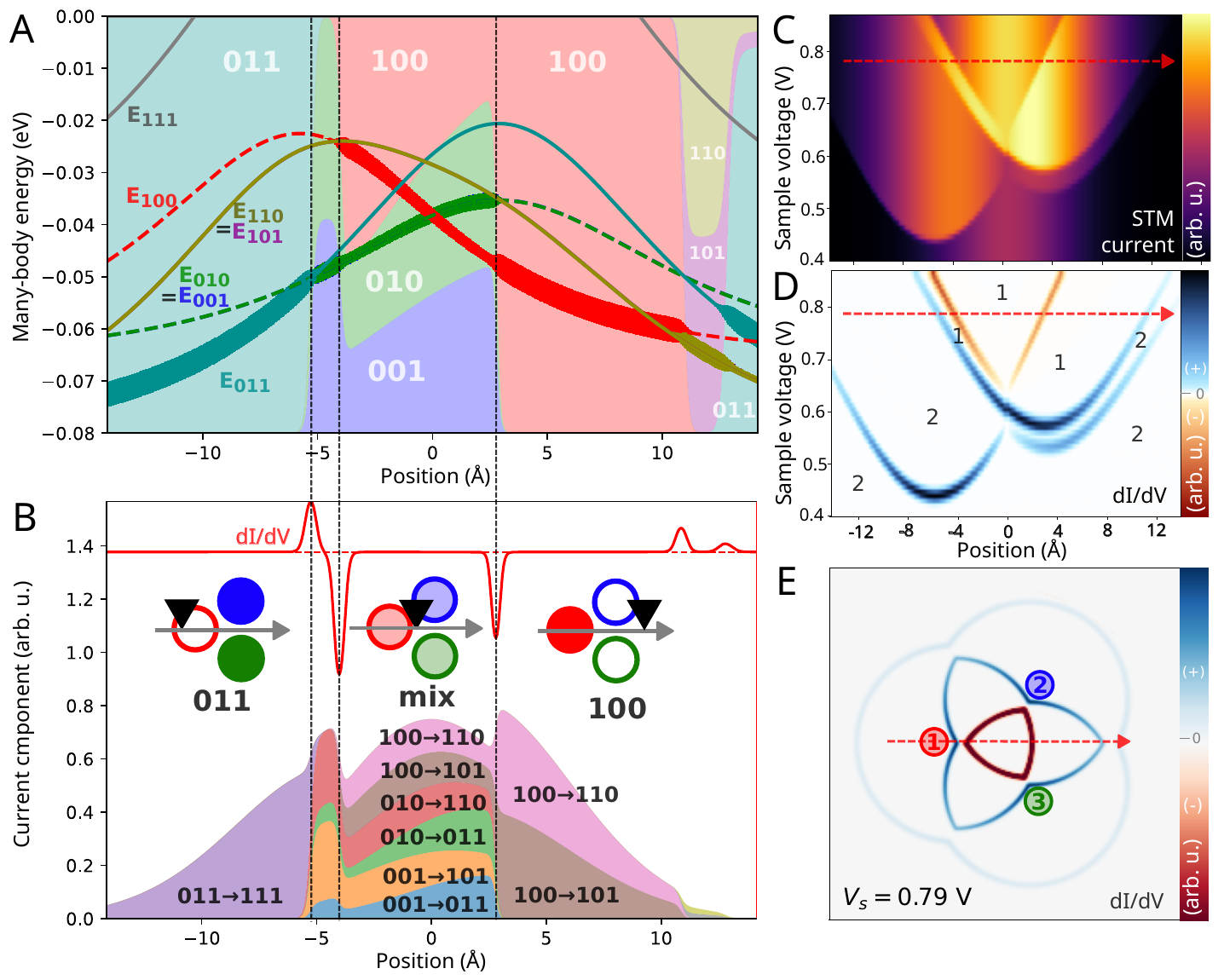}
    \caption{\textbf{Simulation of the discharge rings.} 
        (\textbf{A}) Many-body state energy (lines) and occupation (boldness of line and stacked-area) calculated at sample voltage $V_\text{s}=0.79$~V. The energy of the empty state defines the zero energy. (\textbf{B}) STM current decomposed to transitions between many-body states and d\textit{I}/d\textit{V} at the same voltage. The three schematic drawings illustrates the character of molecular orbital occupancy for given position of the scanning tip. (\textbf{C}) Two-dimensional STM current map with cut at sample voltage $V_\text{s}=0.79$~V depicted by red hashed arrow. (\textbf{D}) Differential conductance maps for the same cut as in panel C. This result can be directly compared to the data in Fig.~\ref{Figure1}F. (\textbf{E}) Spatial map of d\textit{I}/d\textit{V} at $V_\text{s}=0.79$~V. Red arrow marks the path along which the data in panels A-D are plotted.
     }
	\label{Figure3}
\end{figure}

\begin{figure}
	\centering
	\includegraphics[width=0.9\textwidth]{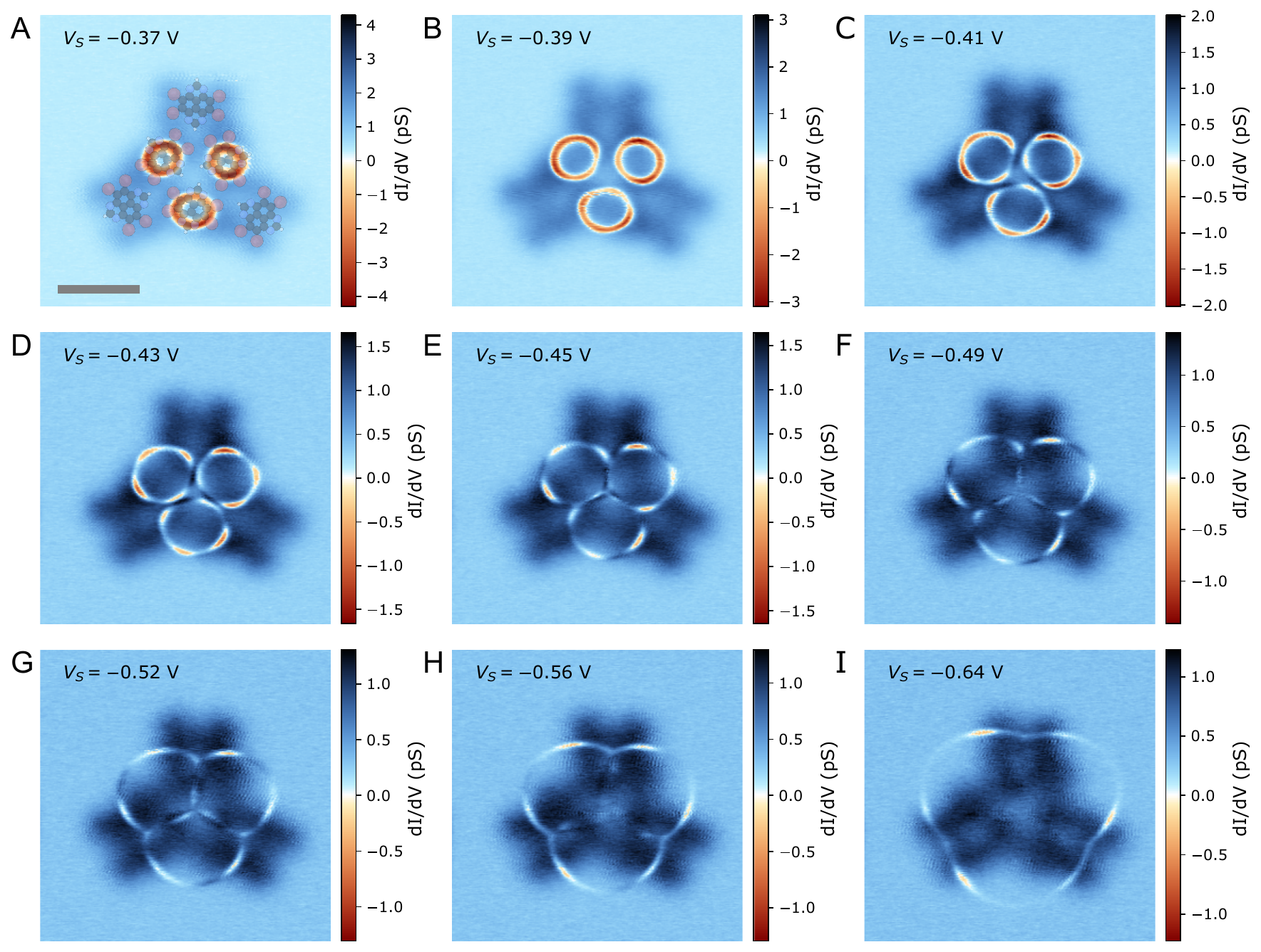}
	\caption{\textbf{Charging rings in the hexamer structure.}
    (\textbf{A-I}) 
     The evolution of charging rings in the hexamer assembly with decreasing sample voltage for $V_\text{s}=-0.37$~V (A), $-0.39$~V (B), $-0.41$~V (C), $-0.43$~V (D), $-0.45$~V (E), $-0.49$~V (F), $-0.52$~V (G), $-0.56$~V (H), and $-0.64$~V (I). Regions of NDC form around the inner molecules with decreasing negative voltage, as a result of the onset of the Coulomb blockade, intersecting at $V_\text{s}=-0.45$~V. Panel A includes an overlaid molecular structure to indicate the positions of the molecules in the hexamer. The scale bar is 1~nm.
    }
	\label{Figure4} 
\end{figure}

\clearpage 

\bibliographystyle{sciencemag}


\section*{Acknowledgments}

V.P., M.\v{Z} and P.H. thank Tom\'{a}\v{s} Novotn\'{y} for many fruitful discussions on the quantum transport theory and the master equation approach.

\paragraph*{Funding:}

This research was supported by the Swiss National Science Foundation (SNSF-grant 200021$\_$204053, 200021$\_$228403, 200020$\_$188445). E.M. and S.-X.L. acknowledge the Sinergia Project funded by the SNSF (CRSII5$\_$213533). E.M. and R.P. acknowledge funding from the European Research Council (ERC) under the European Union's Horizon 2020 research and innovation programme (ULTRADISS grant agreement No 834402). The Swiss Nanoscience Institute (SNI)
is gratefully acknowledged and supports as a part of NCCR SPIN, a National Centre of Competence (or Excellence) in Research, funded by the Swiss National Science Foundation (grant number 51NF40-180604). We gratefully acknowledge the Werner Siemens Stiftung (WSS) for supporting the WSS Research Centre for Molecular Quantum Systems (molQ). C.L. acknowledges the Georg H. Endress Foundation for financial support. M.\v{Z}. and V.P. acknowledge support from Grant No. 23-05263K of the Czech Science Foundation and the EU COST action CA21144 SUPERQUMAP. P.H. acknowledges support from Grant No. 22-06008M of the Czech Science Foundation. Computational resources were provided by the e-INFRA CZ project (ID:90254), supported by the Ministry of Education, Youth and Sports of the Czech Republic.

\paragraph*{Author contributions:}
E.M., R.P., V.P., and C.L.\ designed the experiments. P.Z., S.-X.L.\ and S.D.\ synthesized the molecule. C.L. performed STM experiments. C. L.,  R.P.\ and E.M.\ analyzed the experimental data. V.P., and M.\v{Z}.\ performed the theoretical calculations. P.H. developed the charging simulating method. C.L., V.P., P.H.\ and M.\v{Z}. wrote the manuscript with contributions from R.P. and E.M. All authors discussed the results and revised the manuscript.

\paragraph*{Competing interests:}
There are no competing interests to declare.

\paragraph*{Data and materials availability:}
All data supporting the conclusions of this study are provided in the main text or supplementary materials. The datasets underlying all figures are available on Zenodo.

\subsection*{Supplementary materials}
Materials and Methods\\
Supplementary Text\\
Figs. S1 to S14\\
References \textit{(38-\arabic{enumiv})}\\ 


\newpage


\renewcommand{\thefigure}{S\arabic{figure}}
\renewcommand{\thetable}{S\arabic{table}}
\renewcommand{\theequation}{S\arabic{equation}}
\renewcommand{\thepage}{S\arabic{page}}
\setcounter{figure}{0}
\setcounter{table}{0}
\setcounter{equation}{0}
\setcounter{page}{1}


\begin{center}
\section*{Supplementary Materials for\\ \scititle}
\author{
    Chao Li$^{1,2\dagger\ast}$,
    Vladislav Pokorn\'y$^{3\dagger\ast}$,
    Prokop Hapala$^{3\dagger}$,
    Martin \v{Z}onda$^{4}$,
    Ping Zhou$^{5}$,
    \and
    Silvio Decurtins$^{5}$, 
    Shi-Xia Liu$^{5\ast}$,
    Fengqi Song$^{1}$,
    R\'emy Pawlak$^{2}$,
    Ernst Meyer$^{2}$
    \\
    \small$^{1}$Institute of Atom Manufacturing, Nanjing University, Suzhou 215163, China \and
    \small$^{2}$Department of Physics, University of Basel, Klingelbergstrasse 82, 4056 Basel, Switzerland.\and
    \small$^{3}$Institute of Physics (FZU), Czech Academy of Sciences, Na Slovance 2, 182 00 Prague 8, Czech Republic.\and
    \small$^{4}$Department of Condensed Matter Physics, Faculty of Mathematics and Physics, Charles University, \\
    \small Ke Karlovu 5, 121 16  Prague 2, Czech Republic.\and
    \small$^{5}$Department of Chemistry, Biochemistry and Pharmaceutical Sciences, W. In\"abnit Laboratory \and
    \small for molecular quantum materials and WSS Research Centre for Molecular Quantum Systems, University of Bern, Freiestrasse 3, 3012 Bern, Switzerland.
    \and
    \\
    \small$^\ast$Corresponding author. Email: chao.li@nju.edu.cn; pokornyv@fzu.cz; shi-xia.liu@unibe.ch;
    \\
    \small$^\dagger$These authors contributed equally to this work.
}
\end{center}

\subsubsection*{This PDF file includes:}
Materials and Methods\\
Supplementary Text\\
Figures S1 to S14\\

\newpage

\subsection*{Materials and Methods}

\subsubsection*{Lateral manipulation}
To manipulate individual TBTAP molecules on the Pb(111) surface, we utilized a lateral manipulation technique that is widely applied for moving molecules across metallic substrates. The procedure began by positioning the STM tip directly above the selected molecule using standard imaging conditions ($I = 100$~pA, $V_\text{s} = 100$~mV). Next, the tip was brought closer to the molecule by adjusting the tunneling conditions to $I = 3$~nA and $V_\text{s} = 3$~mV, with the feedback loop turned off. While maintaining this configuration, the tip was laterally translated at a velocity of 200 pm/s, effectively dragging the molecule toward the desired site. Once the final position was reached, the feedback loop was re-engaged, and the initial scanning parameters were restored ($I = 100$~pA, $V_\text{s} = 100$~mV). 

\subsubsection*{STM/STS experiments}
STM and STS experiments were performed with a low-temperature ($2.6$~K) Joule-Thomson STM/AFM microscope (purchased from Omicron GmbH) in ultra-high vacuum (UHV) of $\approx$ $10^{-10}$ mbar operated with Nanonis RC5e electronics. Differential conductance (d\textit{I}/d\textit{V}) spectra were recorded with an internal Nanonis lock-in amplifier using modulation amplitudes indicated in the figure captions. The measurements were performed with a pure Pb tip to enhance the energy resolution beyond the thermal limit.

\subsubsection*{DFT calculations}
The density functional theory was used to find the position of a TBTAP trimer adsorbed onto Pb(111) surface. Calculations were carried out using Turbomole 7.5.1\cite{turbomole} \texttt{ridft} code utilizing the def2-SV(P) (double-zeta) basis set and B3-LYP exchange-correlation functional with DFT-D3 (Becke-Johnson) dispersion correction. The Pb surface was modeled as a slab of three atomic layers. The system was charged by three electrons to simulate charge transfer to the molecules. 

\subsubsection*{Deconvolution of the YSR tunneling spectra }
A deconvolution procedure was utilized to obtain the low-energy, YSR spectra from the d\textit{I}/d\textit{V} data. We used the maximum entropy method implemented in a modified \texttt{ana\_cont} package~\cite{Kaufmann-2023}. Additional details on the deconvolution procedure are presented in~\cite{Li-2025}. A flat default model of the same width as the input data was used. The optimal value of the hyperparameter $\alpha$ was obtained using the \textit{chi2kink} method\cite{Bergeron-2016}. The scanning tip parameters were fitted by a two-step procedure. First, the sum of the gap widths of the tip and the sample $\Delta_\text{t}+\Delta_\text{s}=2.62$~meV was fitted from the tunneling spectra measured on a bare Pb surface. Then, the difference between the gap widths $\Delta_\text{t}-\Delta_\text{s}$ was fitted from the positions of the thermal images, which are well pronounced in the tunneling spectra due to the higher experimental temperature $T_\text{exp}=2.6$~K. The results read $\Delta_\text{t}=1.35$~meV and $\Delta_\text{s}=1.27$~meV. The value of the broadening parameter obtained by the initial fit reads $\gamma_\text{s}=\gamma_\text{t}=0.03$~meV. The modified \texttt{ana\_cont} code is available from authors upon request.

\subsubsection*{NRG calculations of the YSR spectra}
The NRG calculations were performed using the open-source NRG Ljubljana package~\cite{zitkoNRGljubljana}. 
Following previous results~\cite{Li-2025}, we fixed the half-bandwidth of the conduction band to $1$~eV (unity in the code) and the superconducting gap to $\Delta = 1.3$~meV. The charging energy was set to $U = 200$~meV, in line with earlier studies of TBTAP molecules on surfaces~\cite{li2023strong}. For reproducibility, we provide here the specific NRG Ljubljana parameters used in our calculations; detailed descriptions of these parameters are available in the package manual.

We focus on the most strongly correlated system with $\zeta = 1$ (see the discussion below Eq.~\eqref{eq:Gammaij}), which enables single-channel calculations. The subgap spectrum and related quantities were obtained using the following settings: $\texttt{lambda} = 2$, \texttt{symtype} = \texttt{SPSU2}, \texttt{keepenergy} = 10, \texttt{keep} = 4000, and \texttt{keepmin} = 4000. The impurity spectral functions were calculated using $z$-averaging with $N_z = 40$ and the modified log-Gaussian kernel broadening (\texttt{smooth} = \texttt{newsc}), allowing for different broadening values within the superconducting gap ($\omega_0 = 3 \times 10^{-5}$) and above it ($\alpha = 0.05$).

\subsubsection*{Transport calculations}
Simulations of the tunneling spectra for the molecular trimer were performed using a code developed by us, based on the PME approach. The implementation of ME was derived from the \texttt{QmeQ} package~\cite{Kirsanskas-2017}, with an integrated model of the electrostatic field of the tip, SPM image simulations, and a graphical user interface. In these calculations, we neglected the superconducting nature of the substrate, as it does not have any effect on the discharge rings. The system was mapped on the three-impurity Anderson model with two metallic leads, the substrate and the scanning tip. The details of the model and the PME approach are described in the Supplementary Text. Some of the results of our code were verified using the \texttt{QmeQ} package, showing perfect agreement. The code is available at\cite{ppafm_photonmap}. We hope this software will support future studies of similar systems by other experimental groups.

\subsection*{Supplementary Text}

\subsubsection*{Superconducting three-impurity Anderson model}
\label{SM:SC-TIAM}
In all our calculations, including the analysis of the YSR states, we represent the TBTAP trimer using the superconducting three-impurity Anderson model (TIAM) with triangular geometry. The Hamiltonian reads
\begin{equation}
  \label{Eq:Ham}
  \mathcal{H}=\mathcal{H}_\text{imp}+\mathcal{H}_{\text{lead}}+\sum_{i=1}^3\mathcal{H}_\text{hyb}^{(i)}.
\end{equation}
The three impurities are described as
\begin{equation}
  \label{Eq:HamDot}
   \mathcal{H}_\text{imp}
  =\sum_{i\sigma}\left[\varepsilon_{i}(\mathbf{r}) n_{i\sigma}+U_in_{i\uparrow}n_{i\downarrow}\right]
  +\sum_{i<j}W_{ij} n_in_j-\sum_{i<j,\sigma}t_{ij}\left(d_{i\sigma}^\dag d_{j\sigma}^\pdag +\text{H.c.}\right).
 \end{equation}
Here $d_{i\sigma}^\dag$ creates an electron with energy $\varepsilon_{i}(\mathbf{r})=\varepsilon_{id}+\nu_i(\mathbf{r})$ and spin $\sigma$ on the impurity $i$, $\varepsilon_{id}$ is the impurity energy level in absence of the scanning tip, $\nu_i(\mathbf{r})$ is the energy shift due to the presence of the scanning tip at position $\mathbf{r}$, $U_i$ is the local, on-site Coulomb interaction, $W_{ij}$ is the inter-site capacitive coupling between impurities $i$ and $j$, $t_{ij}$ is the direct hopping between impurity orbitals and
$n_{i\sigma}=d_{i\sigma}^\dag d_{i\sigma}^{\phantom{\dag}}$ and $n_i=n_{i\uparrow}+n_{i\downarrow}$ are the particle number operators.
The superconducting substrate (s) and the scanning tip (t) are described by
\begin{equation}
  \label{Eq:HamLeads}
  \mathcal{H}_{\text{lead}}=
  \sum_{\alpha\mathbf{k}\sigma}\varepsilon^\pdag_{\alpha\mathbf{k}}
  c_{\alpha\mathbf{k}\sigma}^\dag c_{\alpha\mathbf{k}\sigma}^\pdag
  - \sum_{\alpha\mathbf{k}}\left(\Delta_\alpha^\pdag c_{\alpha\mathbf{k}\uparrow}^\dag c_{\alpha\mathbf{-k}\downarrow}^\dag+\text{H.c.}\right),
\end{equation}
where $c_{\alpha\mathbf{k}\sigma}^\dag$ creates an electron in lead $\alpha=s,t$ with energy $\varepsilon_{\alpha\mathbf{k}}$ and spin $\sigma$ and $\Delta_\alpha$ is the superconducting order parameter.

The coupling between the impurities and the substrate and the scanning tip is described by
\begin{equation}
  \label{Eq:HamHyb}
  \mathcal{H}_\text{hyb}^{(i)}=
  \sum_{\mathbf{k}\sigma}\left(t_{\text{s}i\mathbf{k}}^\pdag c_{\text{s}\mathbf{k}\sigma}^\dag d_{i\sigma}^\pdag
  +t_{\text{t}i\mathbf{k}}^\pdag(\mathbf{r}) c_{\text{t}\mathbf{k}\sigma}^\dag d_{i\sigma}^\pdag+\text{H.c.}\right),
\end{equation}
where $t_{\alpha i\mathbf{k}}$ is the tunneling matrix element between lead $\alpha=$s,t and impurity $i=1,2,3$. 

In all our calculations, we make the following assumptions: (1) Site-independent value of the capacitive coupling $W_{ij}=W<U$ and hopping $t_{ij}=t$. (2) Same superconducting gap of the scanning tip and the substrate $\Delta_\text{s}=\Delta_\text{t}\equiv\Delta$. (3) Same value of the coupling to the substrate, $t_{\text{s}i\mathbf{k}}=t_{\text{s}\mathbf{k}}$ while $t_{\text{t}i\mathbf{k}}$ depends on the actual position of the scanning tip, as explained below. (4) tunneling rates
\begin{equation}
\Gamma_{\alpha ij}(\omega)=\pi\sum_{\mathbf{k}}t^*_{\alpha i\mathbf{k}}t^{\phantom{*}}_{\alpha j\mathbf{k}}\delta(\omega-\varepsilon^\pdag_{\alpha\mathbf{k}})\equiv\Gamma_{\alpha ij}
\label{eq:Gammaij}
\end{equation}
are energy-independent. From that we can define momentum-independent tunnel matrix elements $t_{\alpha i}=\sqrt{\Gamma_{\alpha ii}/(\pi \rho_{0\alpha})}$, where $\rho_{0\alpha}$ is the Fermi-energy value of the density of the states (DOS) of the lead $\alpha$, which we consider to be constant, $\rho_{\alpha}(\omega)=\Theta(D_\alpha^2-\omega^2)/(2D_\alpha)$, where $D_\alpha$ is the half-bandwidth. This leads to $\rho_{0\alpha}=1/(2D_\alpha)$. We denote the local tunneling rates $\Gamma_{\text{s}ii}\equiv\Gamma_{\text{s}}$ and $\Gamma_{\text{t}ii}\equiv\Gamma_{\text{t}i}$. (5) For the non-local ($i\neq j$) hybridization terms we assume that $\Gamma_{\text{t}ij}=0$ and $\Gamma_{\text{s}ij}=\zeta\Gamma_{\text{s}}$, where $0\leq\zeta\leq 1$ is a parameter that indirectly encodes the distance between the impurities. (6) Both leads are thermalized according to a Fermi–Dirac distribution 
$f_\alpha(\omega)=\big(\exp[(\omega-\mu_\alpha)/k_BT_\text{exp}]+1\big)^{-1}$, 
where $\mu_\alpha$ is the chemical potential of lead $\alpha$ and $T_\text{exp}$ is the experimental temperature.

\subsubsection*{Tip electrostatics}
The influence of the electrostatic field of the tip on molecular energy levels $\varepsilon_{i}(\mathbf{r})$ is the key ingredient in simulating the charging rings. The challenge is that the exact shape of the electrostatic field in STM experiments is unknown, as it depends on the uncontrollable geometry of the tip. A variety of models for AFM/STM tips exist in the literature (e.g., spherical, conical, or pyramidal shapes) \cite{Argento1996,Baier2012,Gao2014,Gross2014}. For this reason, we developed our own electrostatic model suitable for d\textit{I}/d\textit{V} simulations of organic molecules on flat substrates. In our model, the nanojunction is treated as a parallel-plate capacitor defined by two conducting planes at heights $z_\text{C}$ and $z_\text{S}$ and biased by a voltage $V_{\text{s}}$. The molecules are placed at $z=0$, with the substrate (i.e. the second capacitor plate) located at $z_\text{S}<0$ below. A single spherical adatom of radius $r_\text{tip}$ is placed at height $z_\text{tip}$ above the molecules that contributes an additional radial Coulomb-like potential. The bottom capacitor plate is grounded, and this surface need not coincide with the top atomic layer of the substrate. Instead, $z_S$ is treated as a fitting parameter to match experimental features (particularly the evolution of the size of the charging rings with increasing $V_\text{s}$). This surface also acts as an electrostatic mirror for the molecular charge and multipole moments. We found that both the capacitor plate and  the electrostatic mirror are essential to reproduce the exact shape of the charging rings. The molecular orbitals interact with the tip field via monopole charge $1e^-$. To account for the shape of the molecule, we also include a quadrupole moment $Q_{xx}$ aligned along the molecular long axis, which is fitted to reproduce the experimentally observed ellipticity of the charging rings. Due to molecular symmetry, the dipole and other quadrupolar components vanish. Both monopole and quadrupole moments are mirrored across the substrate at $z = -2z_S$.

\subsubsection*{Tip coupling}
Tunneling between the tip and the sample is, naturally, an essential component of STM/STS simulations. In typical quasi-equilibrium situations, when the molecule is strongly coupled to the substrate and its energy levels are filled up to the substrate's Fermi level, the tunneling current $I$ can be calculated using the Fermi golden rule,
\begin{equation}
I(V) \propto \int_V \sum_i |t_{\text{t}i}|^2 \left[f_\text{t}(0) - f_\text{s}(V)\right],
\end{equation}
where $f_\text{t}$ and $f_\text{s}$ are the Fermi functions of the tip and substrate, respectively, and
$t_{\text{t}i}=\langle\psi_{i}|H'|\psi_\text{t}\rangle$ is the tunneling matrix element between the molecular orbitals $\psi_{i}$ (within the tunneling window) and tip states $\psi_\text{t}$. The perturbation $H'$ originates from the potential of one subsystem acting on the other. In practice, it is often approximated as constant, simplifying the tunneling matrix to an overlap, $t_{\text{t}i}\approx C\langle\psi_{i}|\psi_\text{t}\rangle$. For atomically sharp tips, the tip wavefunction $\psi_\text{t}$ is usually modeled as an $s$-orbital. More sophisticated models are generally unfeasible due to limited knowledge about the atomic structure of the tip.

In previous work, we implemented the \texttt{PPSTM} code~\cite{Krejci-2017} for quasi-equilibrium STM/STS simulations, where tunneling matrix elements were evaluated using Chen's rules in a atomic basis~\cite{Chen-1990}. Later, we simplified and optimized this scheme using 2D or 3D convolution of molecular orbitals projected onto a grid, which was used for simulations of the light-STM technique with the \texttt{PPAFM} code~\cite{Dolezal-2021,Oinonen-2024}. This approach leverages the convolution theorem and fast Fourier transform (FFT) to efficiently compute entire 2D images or 3D stacks in one pass. In the present molecular system, the situation is more complex because the system is out of equilibrium and the occupancy of molecular states must determined by solving the master equations. However, tunneling rates into and out of the molecular states, which serve as input parameters to the ME, are still calculated using the same tunneling matrix elements based on the Fermi golden rule assuming s-orbital symmetry of the tip wavefunction $\psi_\text{t}$.

We found that orbital symmetry, reflected in the elements of the tunneling matrix, is essential to reproduce intensity modulation (four lobes separated by nodal planes) and the chiral features of the d\textit{I(x,y)}/d\textit{V} maps of the TBTAP trimer. This approach was used to obtain results in Fig.~\ref{Figure2}. However, for simplicity, most of our simulations were performed using a basic model of exponentially decaying hopping, $t_{\text{t}i}=\exp(-\beta|\mathbf{r}_i-\mathbf{r}_\text{t}|)$, where $\mathbf{r}_i$ is the position of the $i$-th molecule (the center of mass of the relevant molecular orbital), $\mathbf{r}_\text{t}$ is the position of the tip apex, and $\beta$ is a tunable parameter. This corresponds to modeling the orbital $\psi_{i}$ as a spherically symmetric $s$-function. Despite its simplicity, this model is sufficient to qualitatively reproduce all the characteristics of the charging rings, including those associated with NDC, as presented in Fig.~\ref{FigSM:maps_dIdV}.

\subsubsection*{Pauli master equations}
We applied the approximate master equation approach~\cite{Goldozian-2016} to the TIAM without superconducting pairing in the Coulomb blockade regime to obtain the current from the scanning tip to the molecular trimer. Here we assume that the time between two tunneling events is the largest time scale, i.e. the sequential tunneling regime. We also assumed infinite value of the local Coulomb interaction $U$ which makes the energy of the doubly occupied states prohibitively large. This assumption largely simplifies the calculations, as it reduces the dimension of the configurational space from 64 to 8. The value of the molecular energy level in the absence of the scanning tip was fixed to $\varepsilon_{id}=-90$~meV as assumed from the data in Fig.~\ref{FigSM:DiffCondLarge}. The actual scanning-tip-position dependent values of of the molecular energy levels $\varepsilon_{i}(\mathbf{r})$ and the tip-sample tunneling amplitudes $t_{\text{t}i}(\mathbf{r})$ were calculated using the approach from previous sections. 

We diagonalized Hamiltonian ~\eqref{Eq:HamDot} to get the many-body eigenstates (note that this step is not necessary in case of $t=0$ model without direct hopping), $\mathcal{H}_\text{imp}=\sum_aE_a|a\rangle\langle a|$ 
and the many-body tunneling amplitudes
$T^\alpha_{ba}=\sum_{i}V_{\alpha i}\langle b|d^\dag_{i}|a\rangle$, $\alpha=s,t$. To obtain the current through the device we use the simplest PME approach, in which the coherences are not included and only state populations $P_\alpha$ (diagonal elements of the reduced density matrix) are considered. The probabilities are given by the kinetic (master) equations~\cite{Goldozian-2016}
\begin{equation}
\begin{aligned}
\frac{\partial P_{b}}{\partial t}&=
\sum_{\alpha,a}\left[P_{a}\Gamma^\alpha_{a\rightarrow b}f(x^\alpha_{a,b})
-P_{b}\Gamma^\alpha_{b\rightarrow a}f(-x^\alpha_{a,b})\right]\\
&+\sum_{\alpha,c}\left[P_{c}\Gamma^\alpha_{c\rightarrow b}f(x^\alpha_{c,b})
-P_{b}\Gamma^\alpha_{b\rightarrow c}f(-x^\alpha_{c,b})\right]
\end{aligned}
\end{equation}
where $x^\alpha_{a,b}=(E_{a}-E_{b}-\mu_\alpha)/k_BT_\alpha$ and
$\Gamma^\alpha_{a\rightarrow b}=2\pi\rho_{0\alpha}T^\alpha_{ba}T^\alpha_{ab}=\Gamma^\alpha_{b\rightarrow a}$. Here we use the notation in which the particle numbers follow $N_a=N_b-1$ and $N_c=N_b+1$. These equations are solved in the stationary limit $\partial P_b/\partial t=0$ with the normalization condition $\sum_b P_b=1$.

The steady state current from lead $\alpha$ to the molecular trimer is then given by
\begin{equation}
I_\alpha=\frac{e}{\hbar}\sum_{ab}\left[P_{a}\Gamma^\alpha_{a\rightarrow b}f(x^\alpha_{a,b})
-P_{b}\Gamma^\alpha_{b\rightarrow a}f(-x^\alpha_{a,b})\right].
\end{equation}

We chose a small value of the substrate coupling strength $\Gamma_\text{s}\sim 0.01$~meV~$\ll k_BT_\text{exp}$ to stay in the sequential tunneling regime and the validity regime of the PME. This value is much smaller than what can be inferred from the experimental data for TBTAP on Pb(111), $\Gamma_{\text{s},\text{exp}}=10-20$~meV~\cite{Li-2025}. This simplification is justified by the fact that the size and shape of the discharge rings do not depend on the value of $\Gamma$, but rather on the electrostatic details governed by the local energy levels $\varepsilon_{i}(\mathbf{r})$ and the capacitive coupling $W$. Only the width of the discharge ring is proportional to $\Gamma_\text{s}$ and $\Gamma_\text{t}$, while in our calculation it is given solely by the temperature $k_BT_\text{exp}$. 

We obtained the parameters of the model by fitting the experimental data plotted in Fig.~\ref{FigSM:maps_dIdV}. 
The fitted parameters of the electrostatic model read $r_{\text{tip}}=0.3$~nm, $z_{\text{tip}}=0.6$~nm, $z_\text{C}=2$~nm and $z_\text{S}=-0.09$~nm. Note that the position of the electrostatic mirror $z_\text{S}$ does not match the distance of the Pb(111) surface from the molecules which is around 0.3~nm. The fitted value of the inter-site Coulomb coupling reads $W=50$~meV, in agreement with the previous fit of the YSR spectra of TBTAP dimers~\cite{Li-2025}. We also compared the result for infinite value of $U$ with the results for $U=200$~meV calculated using the \texttt{QmeQ} package~\cite{Kirsanskas-2017}. The comparison plotted in Fig.~\ref{FigSM:Udep} shows little or no differences, which justifies the previous assumption. The best fit of the experimental data was obtained for the zero value of inter-site hopping $t$. This further simplifies the analysis, as in the absence of direct hopping, the eigenstates of $\mathcal{H}_\text{imp}$ are just the basis states that we label
\begin{equation}
|000\rangle,\quad|100\rangle,\quad|010\rangle,\quad|001\rangle,
\quad|011\rangle,\quad|101\rangle,\quad|110\rangle,\quad|111\rangle,
\end{equation}
where $|100\rangle=d^\dag_1|000\rangle$, $|110\rangle=d^\dag_2 d^\dag_1|000\rangle$ etc.

The energies of the states for sample voltage $V_\text{s}=0$ read 0, $\varepsilon_{id}$, $2\varepsilon_{id}+W$ and $3\varepsilon_{id}+3W$ for empty, singly occupied, doubly occupied and triply occupied states, respectively. From this we obtain the value of the capacitive coupling $W$ at which the ground state (GS) changes from triply to doubly occupied, $W=-\varepsilon_{id}/2$. For $W=50$~meV and $\varepsilon_{id}=-90$~meV, the GS is doubly occupied, as noted in Figs.~\ref{Figure1}E-F and Fig.~\ref{Figure3}D. The many-body energies of the isolated cluster for two values of the sample voltage $V_\text{s}=0.51$~V and $0.79$~V, which also include the effects of the quadrupole moment, are plotted in Fig.~\ref{FigSM:mbe}. Here, the energy of the empty state $|000\rangle$ marks the energy zero. The ground state for $V_\text{s}=0.51$~V is always doubly occupied, illustrating the situation where crossing the discharge ring is not connected with a change of the total charge, but rather a redistribution of the charge inside the cluster. The triply occupied state always lies at higher energies.

\subsubsection*{YSR spectroscopy}
The spinful character of the TBTAP$^{\bullet-}$ anion deposited on the superconducting surface gives rise to distinct YSR features in the spectra measured at low energies $eV_\text{s}\sim\Delta_\text{s}$, revealing additional details about the charge state of the TBTAP assemblies. The measurements were performed with a pure Pb tip to enhance the energy resolution beyond the thermal limit. As a result, the tunneling spectra are a convolution of the tip and sample DOS. A deconvolution procedure based on the maximum entropy method (see the Methods section) was used to obtain the sample DOS. The results for three positions of the scanning tip presented in Fig.~\ref{FigSM:YSR}A-F show two pairs of well-developed YSR peaks at sample voltages $V_\text{s}=\pm 1.85$~mV and $\pm 1.95$~mV that correspond to YSR energies $E_{YSR}=eV_\text{s}-\Delta_\text{t}=0.50$ and $0.60$~meV. The faint features observed at $V_\text{s}=\pm 0.75$~mV are thermal replicas of the YSR peaks that appear due to the finite experimental temperature $T_\text{exp}=2.6$~K. The spatial map of d\textit{I}/d\textit{V} at $V_\text{s}=\pm 1.8$~mV in Fig.~\ref{FigSM:YSR}I reveals a distinct chiral pattern of the YSR states, similar to the structures observed in the high-energy NDC features presented in Fig.~\ref{Figure2}.

We used the superconducting TIAM to describe the properties of the YSR states. The model was solved using NRG (see the Methods section), using similar parameters as for the simulation of the discharge rings, $U=200$~meV, $W=50$~meV and $t=0$. We neglect the coupling to the scanning tip $\Gamma_\text{t}$ and use a realistic value of the coupling to the substrate, $\Gamma_{\text{s}i}=\Gamma_i=20$~meV~\cite{Li-2025} with $\zeta=1$.  The effect of the scanning tip is discussed later. We also assume the same local energy on each site, $\varepsilon_{id}=\varepsilon$.

Fig.~\ref{FigSM:YSR2}A shows the evolution of the sub-gap many-body spectrum calculated using NRG with respect to the GS energy $E_0$, as a function of the on-site energy $\varepsilon$. The data show a pair of quantum phase transitions, first at $\varepsilon=-58.5$~meV between the singlet and the doublet GS, and a second at $\varepsilon=-111.5$~meV between the doublet and the triplet GS (compare to the exact result for vanishing $\Gamma$, which predicts the transition at $\varepsilon=-2W=-100$~meV). The average occupation of the trimer is plotted in Fig.~\ref{FigSM:YSR2}B. The result is in agreement with the PME calculation that also predicts the charge $-2e$ for small values of the sample voltage $V_\text{s}$ and
realistic values of the local energy $\varepsilon\sim-100$~meV. The YSR states in this region correspond to doublet-singlet transitions (orange dots in panel A). In the experiment, the YSR states are split by approx. 0.1~meV. Such a splitting can be the result of an asymmetry in the trimer structure, induced by the attractive force between the scanning tip and the molecules, which decreases the surface coupling. We simulate the effect of the tip by modifying the coupling $\Gamma_1$ of one molecule. The results in Fig.~\ref{FigSM:YSR2}C indeed show splitting of the many-body energies, which are otherwise degenerate pairs of doublets and singlets. As the YSR states correspond to the allowed transitions between the many-body states (e.g., doublet-singlet), this effect is a possible cause of the splitting observed in the experiment. Fig.~\ref{FigSM:YSR}G shows the DOS calculated using NRG for two cases, symmetric case with $\Gamma_1=\Gamma_2=\Gamma_3=20$~meV (red line) and slightly asymmetric case for $\Gamma_1=19.5$~meV (blue line). The small asymmetry in the coupling leads to splitting of the YSR states as predicted, which is in agreement with the experimental result.

YSR spectroscopy was also performed on a hexamer assembly, Fig.~\ref{FigSM:hedidv}D, revealing a YSR peak on the outer three molecules. No YSR features were observed on the inner three molecules, except a faint peak that can be attributed to the tail of a YSR state from an outer molecule. This provides further proof of the loss of spinful radical nature of the inner molecules, as discussed in the main text.

\newpage
\begin{figure} 
	\centering
	\includegraphics[width=0.9\textwidth]{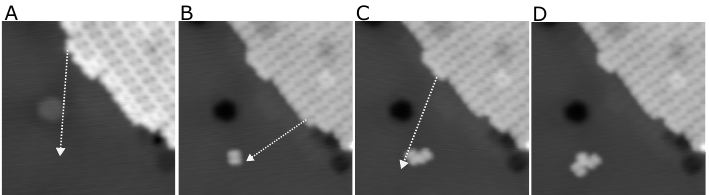}
	\caption{\textbf{Lateral manipulation of TBTAP molecules}. (\textbf{A}–\textbf{D}) Sequential construction of a trimer by laterally relocating individual TBTAP molecules from a molecular island. All images cover an area of $15 \times 15~\text{nm}^2$. Imaging parameters: (\textbf{A}) $V_\text{s} = 1$~V, $I = 60$~pA; (\textbf{B}–\textbf{D}) $V_\text{s} = 80$~mV, $I = 100$~pA.  }
	\label{FigSM:lateral}
\end{figure}

\begin{figure}
	\centering
	\includegraphics[width=0.9\textwidth]{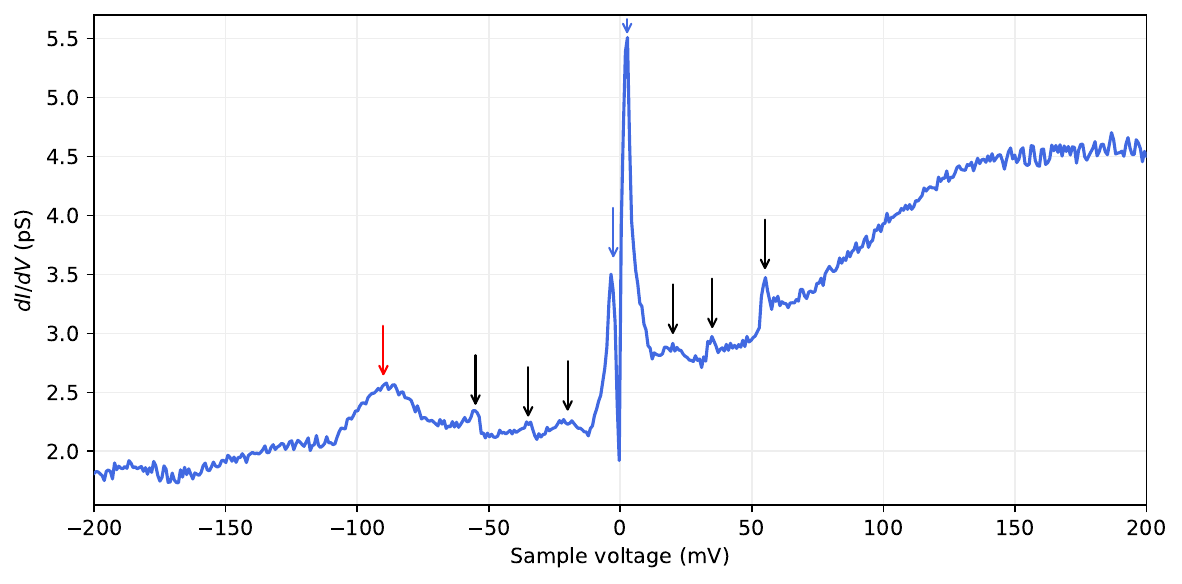}
	\caption{\textbf{Differential conductance spectra of a TBTAP trimer measured on -200 to 200~mV}. The dominant features at $\pm2.7$~mV (blue arrows) mark the coherence peaks at the edges of the superconducting gap. The conductance peaks at $\pm20$, $\pm35$ and $\pm55$~mV (black arrows) can be attributed to vibrations. The broader peak at $-90$~mV is attributable to the singly/lowest occupied molecular orbital (SOMO/LUMO). Note that as the data were measured using superconducting tip, all spectral features are shifted away from the Fermi energy energy by the tip gap $\Delta_t=1.35$~meV. Spectra parameters: $V_{\text{s}}=200$~mV, $A_{\text{mod}}=3$~mV and $f=613$~Hz.}
	\label{FigSM:DiffCondLarge}
\end{figure}

\begin{figure}
	\centering
	\includegraphics[width=0.8\textwidth]{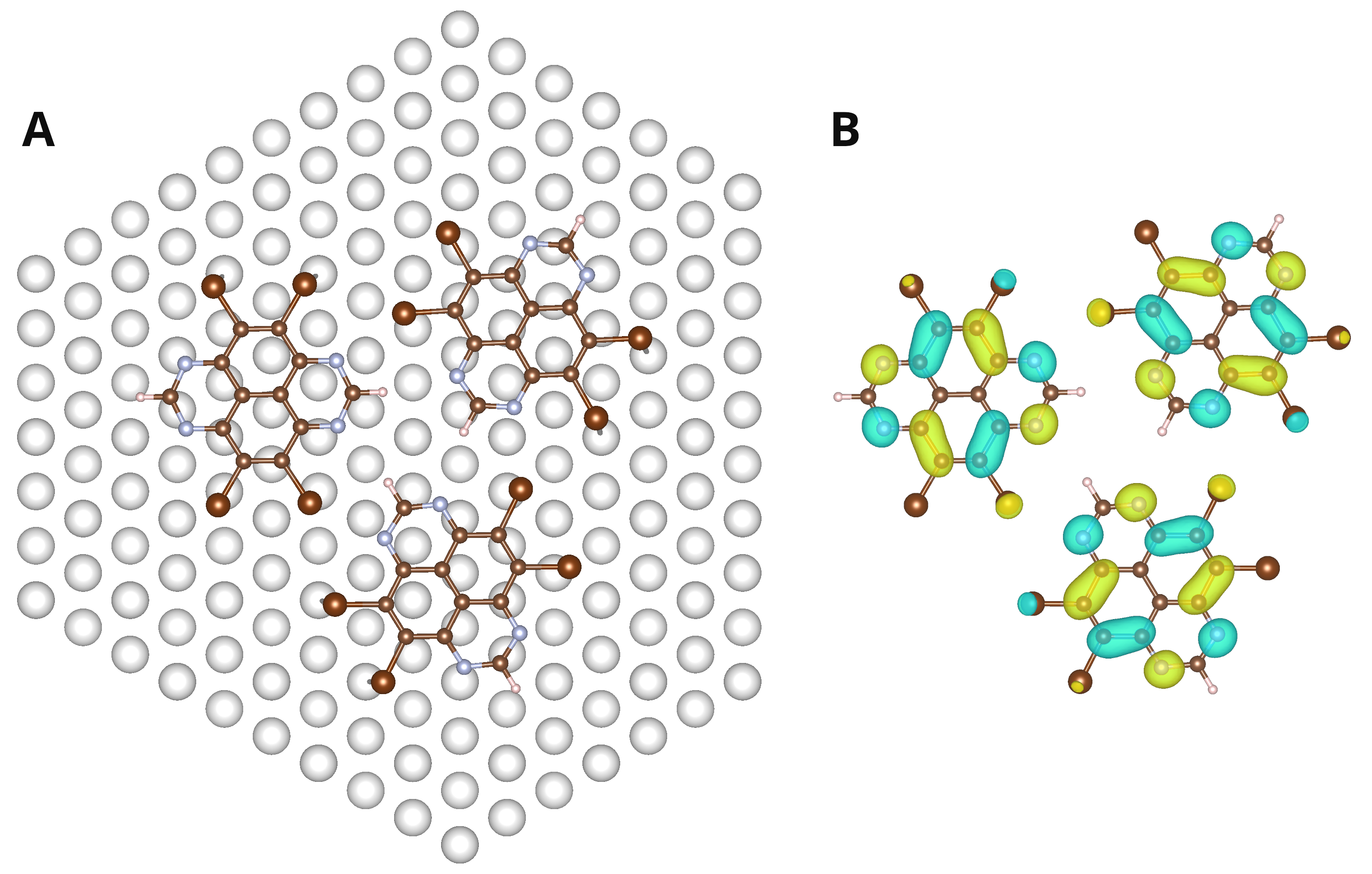}
	\caption{\textbf{DFT result on a TBTAP trimer.}
    (\textbf{A}) Equilibrium position of a TBTAP trimer on Pb(111), that shows the chiral pattern as a result of the incommensurate size of the molecule with respect to the surface lattice parameters. Molecules form an equilateral triangle with $1$~nm distance between the centers of mass of the molecules. The molecules' shorter axes are rotated by $16^\circ$ with respect to the center of the triangle. (\textbf{B}) Isosurface plot of the SOMO orbital of a TBTAP trimer charged by 3 electrons.}
	\label{FigSM:DFT} 
\end{figure}

\begin{figure}
	\centering
	\includegraphics[width=1.0\textwidth]{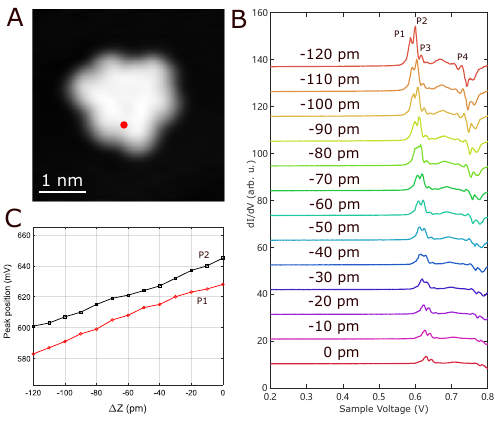}
	\caption{\textbf{Evolution of STS peaks during tip approach}. (\textbf{A} and \textbf{B}) A set of high-energy STS spectra were acquired at the center of a molecule (indicated by the red dot) as the tip-molecule separation was gradually reduced. The spectra are vertically offset for clarity. Distinct spectral features are marked and labeled as P1 to P4. (\textbf{C}) The variation in the energy positions of peaks P1 and P2 is plotted as a function of the tip approach distance. The STM image cover an area of $4 \times 4~\mathrm{nm}^2$. Imaging parameters: $V_\text{s} = 100$~mV, $I = 100$~pA. }
	\label{FigSM:differentheight} 
\end{figure}

\begin{figure}
	\centering
	\includegraphics[width=1.0\textwidth]{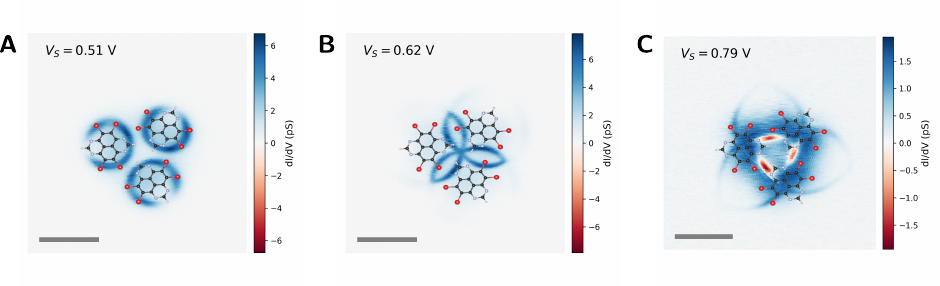}
	\caption{\textbf{Overlay of the molecular structure and discharging rings.}. Approximate orientation of the molecules as obtained from DFT calculation with respect to the discharge rings for three values of the sample voltage $V_\text{s}=0.51$~V (\textbf{A}), $0.62$~V (\textbf{B}) and $0.79$~V (\textbf{C}). The size of the molecules was slightly scaled along the vertical axis to fit the charging rings. This small discrepancy is probably a result of an imperfect calibration of the experimental setup.
    }
	\label{FigSM:superposition} 
\end{figure}

\begin{figure}
    \centering
    \includegraphics[width=1.0\textwidth]{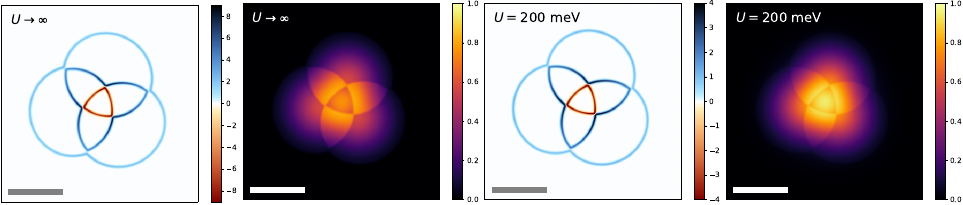}
    \caption{\textbf{Effect of finite charging energy.} Comparison of the Pauli ME results on d\textit{I}/d\textit{V} and current (both in arb. units) for $V_\text{s}=0.72$~V for infinite value of the on-site Coulomb interaction $U$, which prohibits double occupancy of a site, with result for $U=200$~meV. The finite value of the on-site Coulomb interaction leads to larger values of the current, but to negligible differences in d\textit{I}/d\textit{V}. Data were calculated using the \texttt{QmeQ} package~\cite{Kirsanskas-2017}. The bar represents 1~nm.
    }
    \label{FigSM:Udep} 
\end{figure}

\begin{figure}
    \centering
    \includegraphics[width=0.75\textwidth]{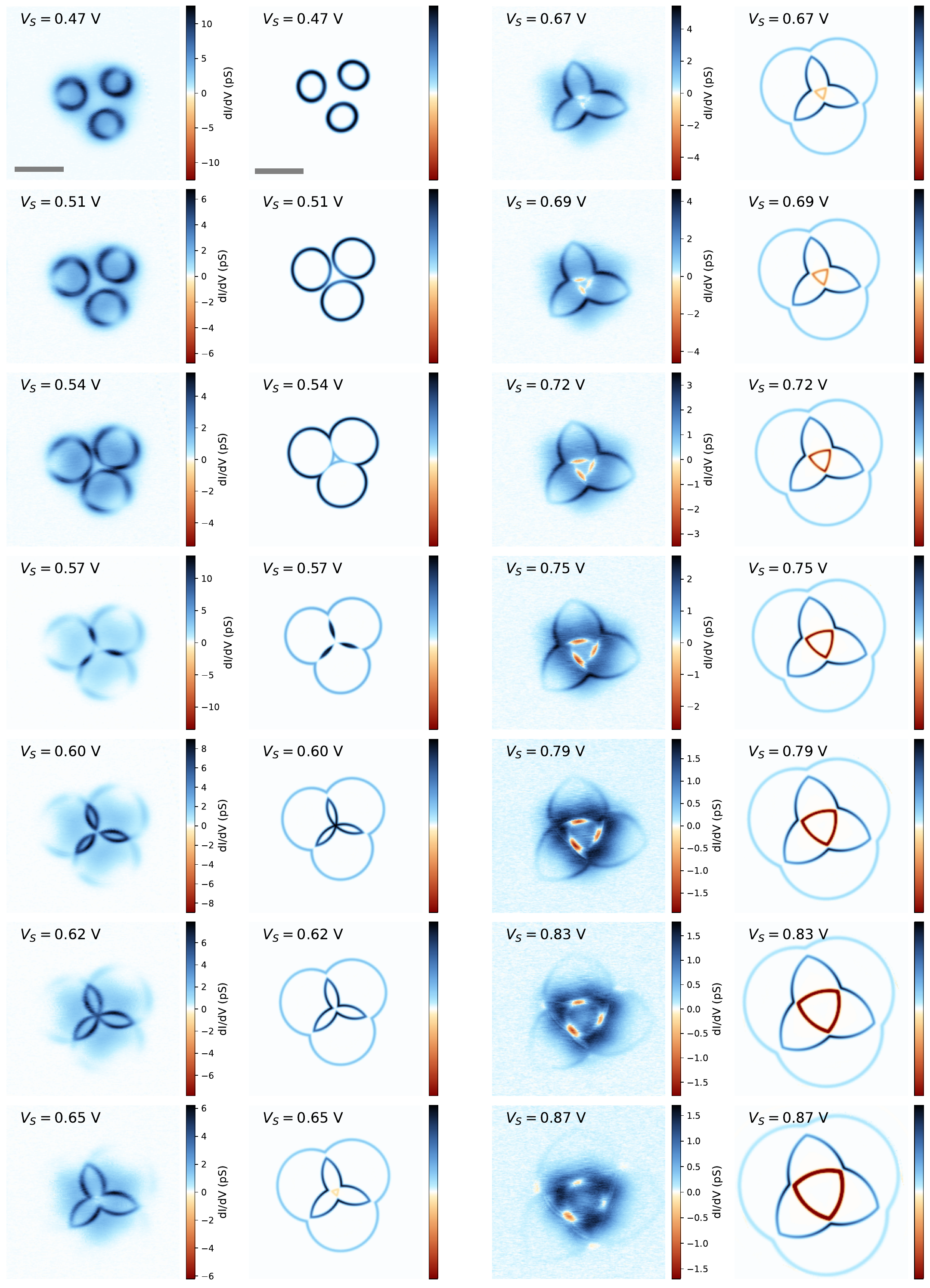}
    \caption{\textbf{Spatial variation of differential conductance in a TBTAP trimer at varying bias voltage - experiment vs. theory.} First and third column: Experimental d\textit{I}/d\textit{V} maps of a TBTAP trimer at different sample voltage $V_\text{s}$. The red color represents regions of NDC. Note the different maxima on the color scales. Second and fourth column: Simulation of the charging rings using the PME approach (in arb. u.). Here, the effects of the TBTAP orbital structure is omitted for clarity. The scale bar is 1~nm.
    }
    \label{FigSM:maps_dIdV}
\end{figure}

\begin{figure}
    \centering
    \includegraphics[width=0.75\textwidth]{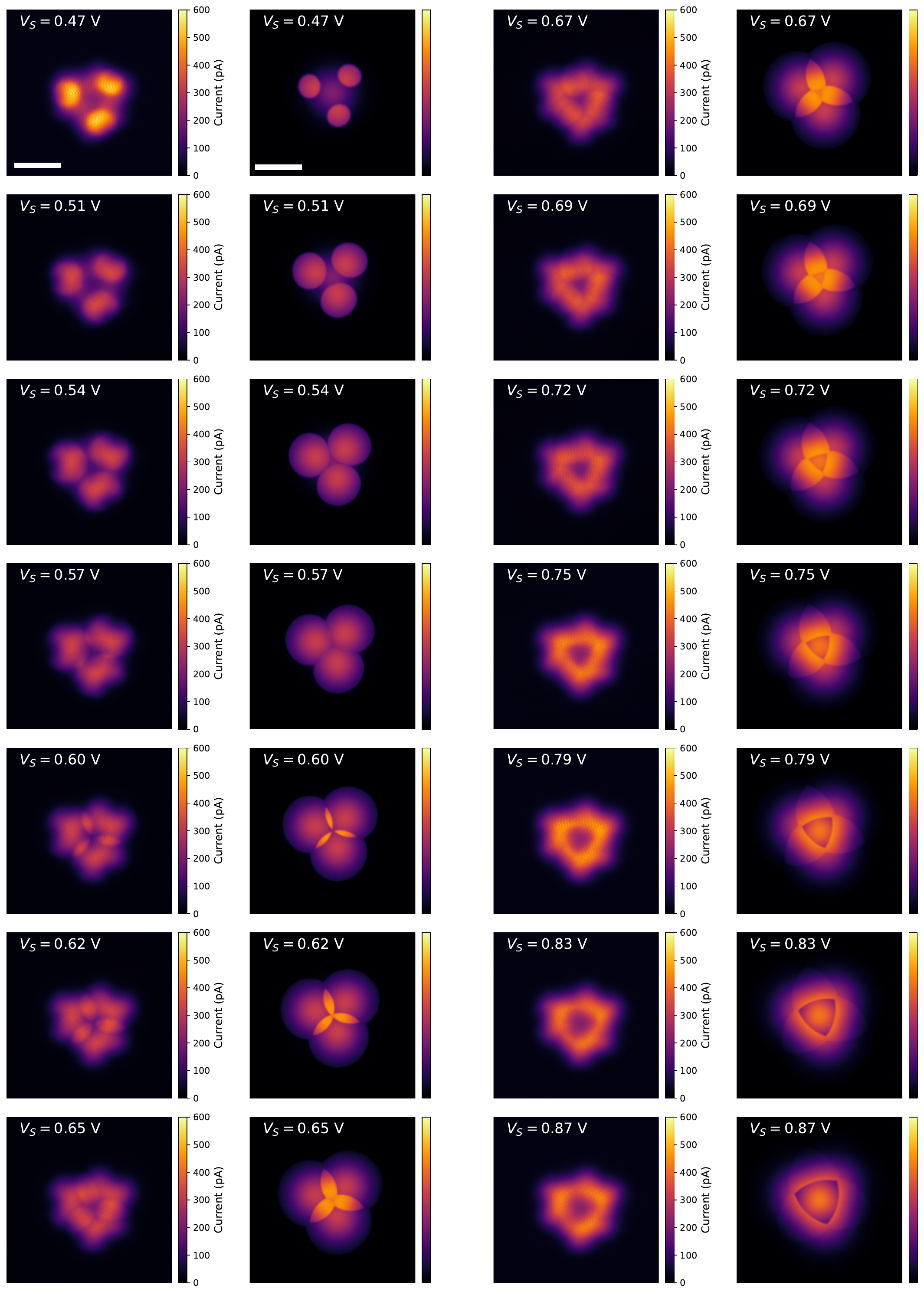}
    \caption{\textbf{Spatial variation of STM current in a TBTAP trimer at varying bias voltage - experiment vs. theory.} First and third column: STM current maps from experiment on a TBTAP trimer at the same sample voltages $V_\text{s}$ as in Fig.~\ref{FigSM:maps_dIdV}. Second and fourth column: STM current maps simulated using PME approach (in arb. u.). The color scales have the same maxima for easier comparison. The scale bar is 1~nm.
    }
    \label{FigSM:maps_current} 
\end{figure}

\begin{figure}
    \centering
    \includegraphics[width=0.9\textwidth]{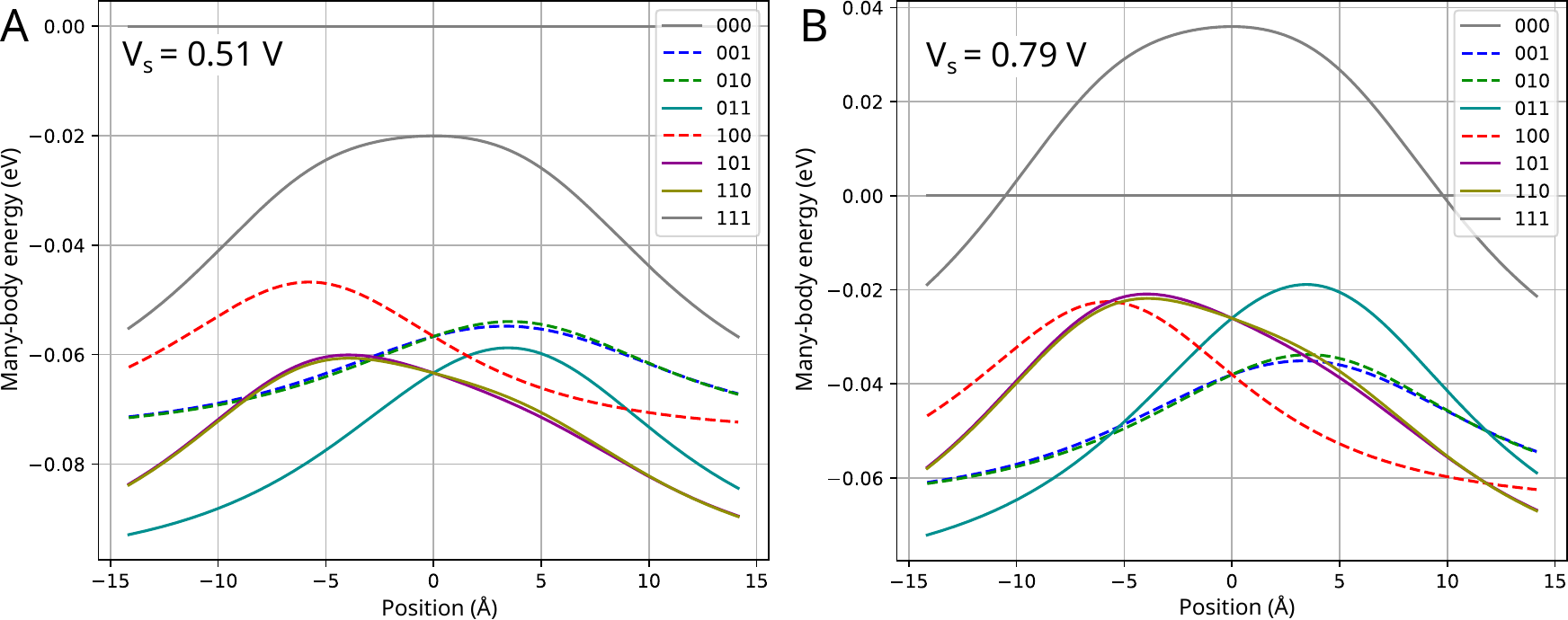}
    \caption{\textbf{Many-body energies of the isolated trimer.} Many-body spectra calculated for sample voltages $V_\text{s}=0.51$~V (\textbf{A}) and $0.79$~V (\textbf{B}) along the same path as in Figs.~\ref{Figure3}A-D, that include the effects of the quadrupole moment. As a result, the singly and doubly occupied states are slightly split, in contrast to the data shown in Fig~\ref{Figure3}A. Results in panel A also show that the ground state for lower voltages is always doubly occupied, as noted in Fig.~\ref{Figure3}D. The triply occupied state lies higher in energy due to the sizable inter-site Coulomb coupling.
    }
    \label{FigSM:mbe}
\end{figure}

\begin{figure}
	\centering
	\includegraphics[width=0.8\textwidth]{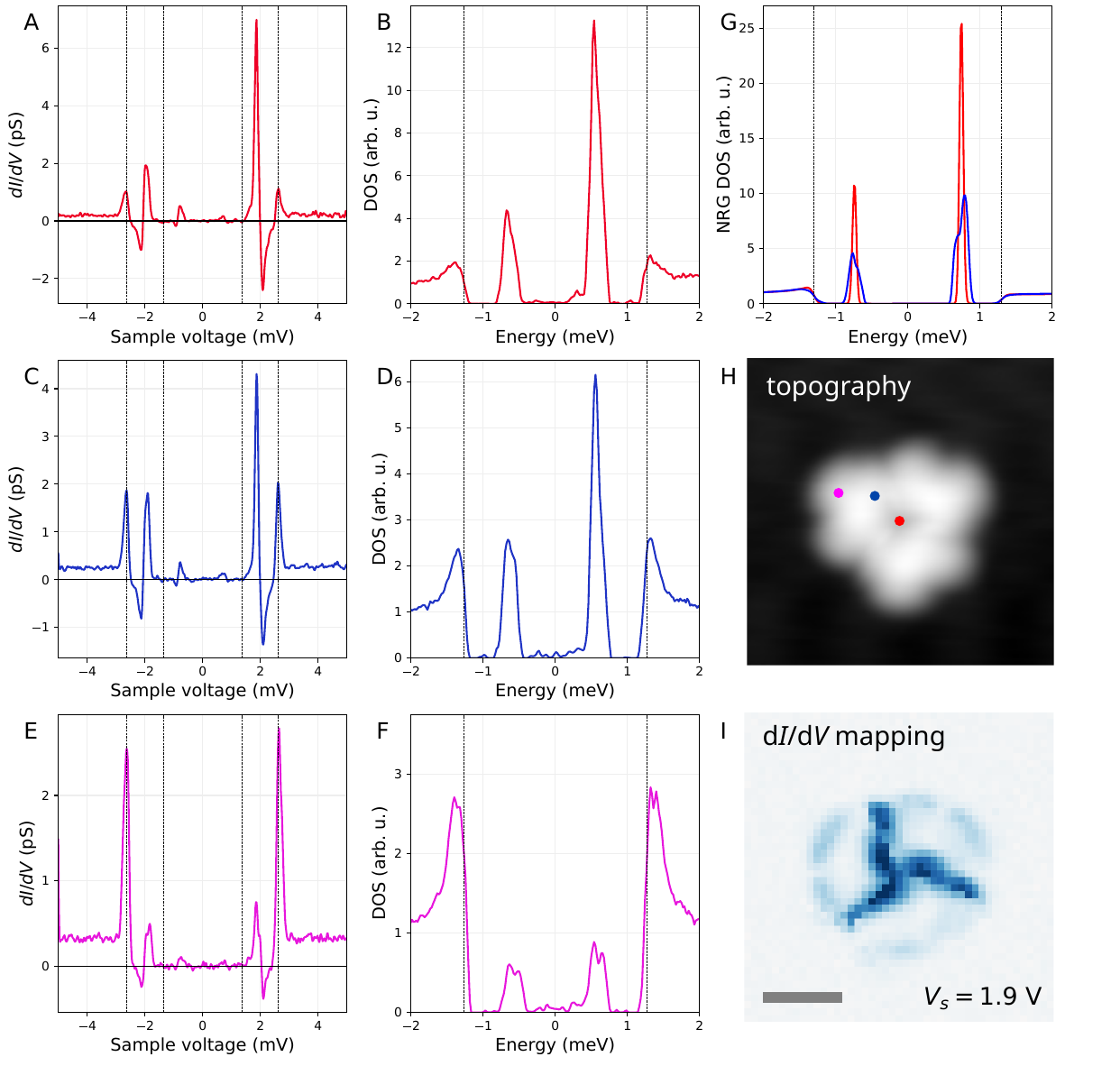}
    \caption{\textbf{YSR spectra of TBTAP trimers}. 
    \textbf{(A–F)} d\textit{I}/d\textit{V} spectra measured at the positions marked by red, blue, and violet dots in panel H and the corresponding surface DOS. In addition to the coherence peaks at $V_\text{s}=\pm 2.7$~mV, two pairs of YSR states appear at $V_\text{s}=\pm 1.85$~mV and $V_\text{s}=\pm 1.95$~mV. Faint features observed at $V_\text{s}=\pm 0.75$~mV are thermal replicas of the YSR peaks. Vertical lines in panel A mark the values of $\pm\Delta_\text{t}$ and $\pm(\Delta_\text{s}+\Delta_\text{t})$. Vertical lines in panel B mark the gap edges $\pm\Delta_\text{s}$.
    \textbf{(G)} NRG result on the YSR spectra of the superconducting TIAM. Model parameters read $\Delta=1.3$~meV, $U=200$~meV, $W=50$~meV, $\varepsilon=-104$~meV, $t=0$ and $\Gamma_2=\Gamma_3=20$~meV. Red line: symmetric case ($\Gamma_1=20$~meV). Blue line: slightly asymmetric case ($\Gamma_1=19.5$~meV) to simulate the effect of the scanning tip. The asymmetry leads to the splitting of the YSR peaks.
    \textbf{(H)} Topographic image of the TBTAP trimer.
    \textbf{(I)} Spatially resolved d\textit{I}/d\textit{V} map acquired at $V_\text{s} = 1.9$~mV, illustrating the chiral nature of the YSR state.
    } 
   \label{FigSM:YSR} 
\end{figure}

\begin{figure}
	\centering
	\includegraphics[width=0.9\textwidth]{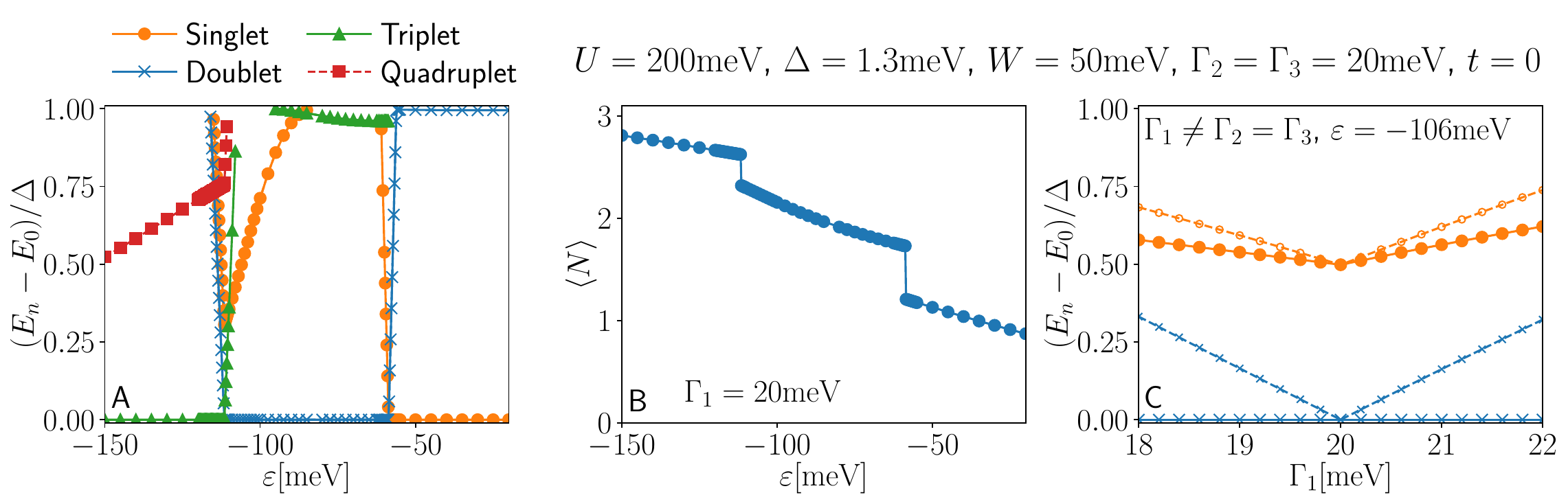}
    \caption{\textbf{Evolution of YSR states in TBTAP trimers}.
    \textbf{(A)} Many-body state energies as functions of the local energy $\varepsilon$ measured with respect to the ground state. The color encodes the state degeneracy. The quantum phase transitions at $\varepsilon=-58.5$~meV and $\varepsilon=-111.5$~meV are marked by the change of the ground state degeneracy. Model parameters read $U=200$~meV, $W=50$~meV, $\Gamma_1=\Gamma_2=\Gamma_3=20$~meV and $t=0$.
    \textbf{(B)} Total charge of the trimer as function of the local energy $\varepsilon$. The total charge for $\varepsilon=-90$~meV is two, in agreement with the PME result.
    \textbf{(C)} The effect of asymmetry induced by varying coupling to the substrate $\Gamma_1$ of one of the molecules. Such variation can be induced by the presence of a scanning tip and leads to a splitting of the YSR energies, in agreement with the experimental result.
    } 
   \label{FigSM:YSR2}
\end{figure}

\begin{figure}
	\centering
	  \includegraphics[width=0.8\textwidth]{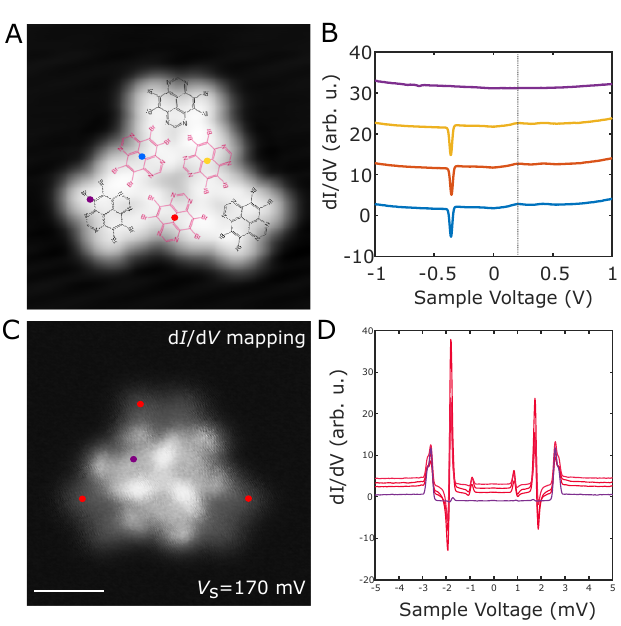}
	  \caption{\textbf{Structure of a TBTAP hexamer}. 
    (\textbf{A}) STM image and structural model of a triangular hexamer constructed from six TBTAP molecules. ($I = 100$ pA, $V_{\text{s}} = -600$ mV) 
    (\textbf{B}) d\textit{I}/d\textit{V} spectra measured on three inner and one outer TBTAP molecules at the positions marked by dots in panel A. Pronounced charging dips are observed between $-400$~mV and $-500$~mV for the inner molecules, while no significant features appear for the outer molecule. In addition to the dips at negative $V_\text{s}$, two resonances are observed at $V_\text{s}=170$ mV and $V_\text{s}=330$~mV, which can be attributed to the LUMO and LUMO+1 energy levels of the inner molecules of the hexamer. 
    (\textbf{C}) Constant-height d\textit{I}/d\textit{V} map acquired at $V_{\text{s}} = 170$ mV, highlighting the spatial variation in electronic structure between the inner and outer TBTAP molecules. 
    (\textbf{D}) YSR spectra measured on one inner and three outer TBTAP molecules at positions marked by dots in panel C. The inner molecule shows no YSR peaks, revealing the loss of its radical nature.}
    \label{FigSM:hedidv} 
\end{figure}

\begin{figure}
	\centering
	\includegraphics[width=1.0\textwidth]{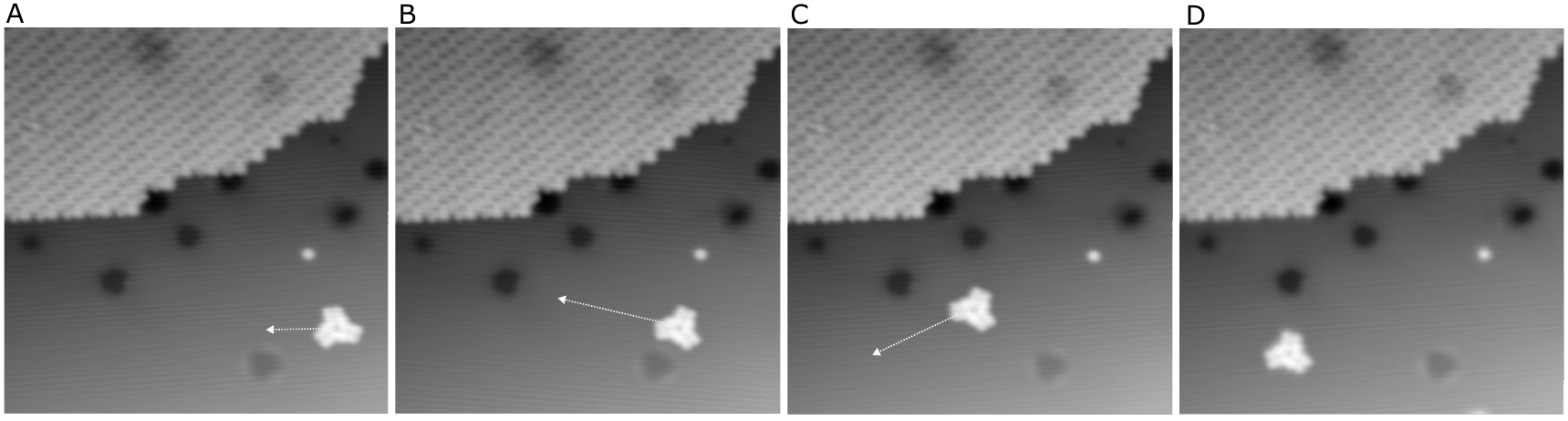}
	\caption{\textbf{Movement of the hexamer assembly using lateral manipulation}. (\textbf{A}–\textbf{D}) Sequential lateral movement of a hexamer structure, composed of six TBTAP molecules, demonstrating its structural integrity. The hexamer is stabilized through a combination of hydrogen and halogen bonding. Images cover an area of $25 \times 25~\mathrm{nm}^2$. Imaging parameters: (\textbf{A}) $V_\text{s} = 60$~mV, $I = 80$~pA; (\textbf{B}–\textbf{D}) $V_\text{s} = 80$~mV, $I = 100$~pA.
    }
	\label{FigSM:hexla} 
\end{figure}

\begin{figure}
	\centering
	\includegraphics[width=1.0\textwidth]{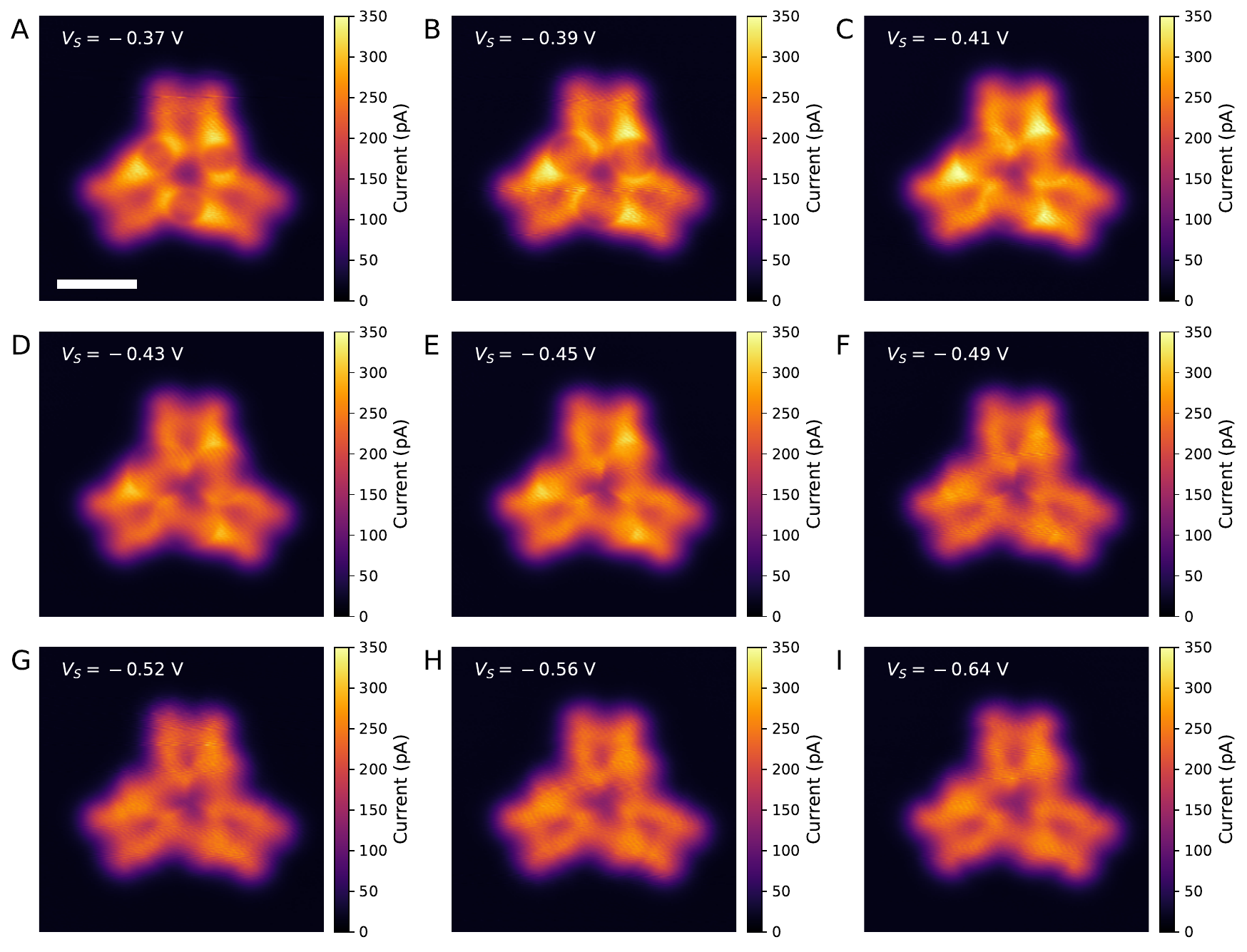}
	\caption{\textbf{Spatial variation of current in a TBTAP hexamer at varying bias voltage}. (\textbf{A}–\textbf{I}) Spatial maps of the absolute value of the current plotted for the same parameters as in Figure~\ref{Figure4} in the main text. The chiral structure of the hexamer is more pronounced than in the d\textit{I}/d\textit{V} maps. The color scales have the same maxima for easier comparison. The scale bar is 1~nm.
    }
	\label{FigSM:hex-current} 
\end{figure}

\clearpage 

\end{document}